\documentclass[published]{JHEP3} 

\JHEP{00(2010)000}

\JHEPspecialurl{http://jhep.sissa.it/JOURNAL/JHEP3.tar.gz}

\usepackage{epsfig,multicol,bbm}

\newcommand{\eqref}[1]{Eq.~(\ref{#1})}

\newcommand{\be}{\begin{equation}}
\newcommand{\ee}{\end{equation}}
\newcommand{\ben}{\begin{equation*}}
\newcommand{\een}{\end{equation*}}
\newcommand{\bea}{\begin{eqnarray}}
\newcommand{\eea}{\end{eqnarray}}
\newcommand{\bean}{\begin{eqnarray*}}
\newcommand{\eean}{\end{eqnarray*}}
\newcommand{\brr}{\begin{array}}
\newcommand{\err}{\end{array}}
\newcommand{\bc}{\begin{center}}
\newcommand{\ec}{\end{center}}

\newcommand{\eg}{\mbox{\it e.g.~}}
\newcommand{\ie}{\mbox{\it i.e.~}}
\newcommand{\cf}{\mbox{\it cf.~}}
\newcommand{\vev}[1]{\mbox{$\langle #1 \rangle $}}

\newcommand{\bk}{{\mathbf k}}

\newcommand{\bp}{{\mathbf p}}

\newcommand{\bx}{{\mathbf x}}

\newcommand{\bq}{{\mathbf q}}

\newcommand{\HH}{{\mathcal H}}

\newcommand{\B}{{\rm B}}

\newcommand{\al}{\alpha}
\renewcommand{\b}{\beta}
\newcommand{\de}{\delta}
\newcommand{\De}{\Delta}
\newcommand{\ep}{\epsilon}
\newcommand{\ga}{\gamma}

\newcommand{\Om}{\Omega}

\newcommand{\Dst}{D^{\rm as}}
\newcommand{\Phist}{\Phi^{\rm as}}
\newcommand{\Psist}{\Psi^{\rm as}}
\newcommand{\Vst}{V^{\rm as}}
\newcommand{\mO}{\mathcal{O}}

\newcommand{\rnu}{15+4R_\nu}

\title{CMB temperature anisotropy at large scales induced by a causal primordial magnetic field}

\author{Camille Bonvin\\
	CEA, IPhT \& CNRS, URA 2306, F-91191 Gif-sur-Yvette, France\\
	E-mail: \email{camille.bonvin@cea.fr}}
\author{Chiara Caprini\\
	CEA, IPhT \& CNRS, URA 2306, F-91191 Gif-sur-Yvette, France\\
	E-mail: \email{chiara.caprini@cea.fr}}

\received{April 8, 2010} 		
\accepted{April 29, 2010}		

\abstract{We present an analytical derivation of the Sachs Wolfe effect sourced by a primordial magnetic field. In order to consistently specify the initial conditions, we assume that the magnetic field is generated by a causal process, namely a first order phase transition in the early universe. As for the topological defects case, we apply the general relativistic junction conditions to match the perturbation variables before and after the phase transition which generates the magnetic field, in such a way that the total energy momentum tensor is conserved across the transition and Einstein's equations are satisfied. We further solve the evolution equations for the metric and fluid perturbations at large scales analytically including neutrinos, and derive the magnetic Sachs Wolfe effect. We find that the relevant contribution to the magnetic Sachs Wolfe effect comes from the metric perturbations at next-to-leading order in the large scale limit. The leading order term is in fact strongly suppressed due to the presence of free-streaming neutrinos. We derive the neutrino compensation effect dynamically and confirm that the magnetic Sachs Wolfe spectrum from a causal magnetic field behaves as $\ell(\ell+1)\,C^\B_\ell \propto \ell^2$ as found in the latest numerical analyses.}

\begin{document} 

\section{Introduction}

The origin of the large scale magnetic fields observed in galaxies and clusters is still unknown: one of the possible explanations is that they have been generated in the primordial universe. A primordial magnetic field of the order of the nanoGauss could leave a detectable imprint in the cosmic microwave background (CMB) anisotropies. This has been analysed in several works: for the scalar mode, see for example \cite{Kahniashvili:2006hy,Yamazaki:2008gr,Giovannini:2007qn,Finelli:2008xh,PFP,Shaw:2009nf,yamafin}. Here we concentrate on the effect that a primordial magnetic field can have on the temperature CMB spectrum at large scales: in particular, we focus on the Sachs Wolfe effect. The motivation is, that conflicting results are present in the literature regarding the $\ell$-dependence of the Sachs Wolfe effect induced at large scales by a primordial magnetic field generated by a phase transition: the analytical analysis of \cite{Kahniashvili:2006hy} found $\ell(\ell+1)\,C^\B_\ell$ scaling as $ \ell^{-1}$ or more negative, and the same result was found in the numerical calculation of \cite{Yamazaki:2008gr}; on the other hand, \cite{Finelli:2008xh,PFP,Shaw:2009nf,yamafin,Caprini:2009vk} found $\ell(\ell+1)\,C^\B_\ell$ scaling as $\ell^2$. 

The aim of this paper is to solve this discrepancy analytically, and we find that the relevant contribution to the Sachs Wolfe effect is the one found in \cite{Finelli:2008xh,PFP,Shaw:2009nf,yamafin,Caprini:2009vk}. Naively, the magnetic field anisotropic stress would induce a CMB spectrum $\ell(\ell+1)\,C^\B_\ell\propto \ell^{-1}$ at large scales, basically due to the fact that the metric perturbation $\Phi$ is proportional to $\Phi \propto \Pi_\B (\HH/k)^2$ at leading order in the large scale expansion $k/\mathcal{H}\ll 1$, where $\Pi_\B$ is the magnetic field anisotropic stress. However, as soon as neutrinos decouple and start to free-stream, they develop a non-zero anisotropic stress which adjusts to compensate the one coming from the magnetic field, see \cite{Kojima:2009gw}. We demonstrate here that this compensation cancels the leading order contribution to the CMB spectrum, and the dominant contribution becomes the one from the next-to-leading order in the $k/\mathcal{H}\ll 1$ expansion, namely $(k/\mathcal{H})^0$, which induces then $\ell(\ell+1)\,C^\B_\ell \propto \ell^2$. A residual $\ell^{-1}$ contribution to the Sachs Wolfe remains: this is the relic effect of the period of time after the magnetic field generation but before the decoupling of neutrinos. However, this contribution is completely unobservable in the CMB today.

In this paper we concentrate on causal magnetic fields, generated at a phase transition in the radiation dominated universe before neutrino decoupling (see for example \cite{phase}). We make this choice in order to be able to consistently set the initial conditions (in a following work we will analyse the inflationary generation case \cite{inprep}). We assume that the magnetic energy momentum tensor is first order in perturbation theory, and that the magnetic field has no background contribution: it is a stochastic primordial magnetic field with spectral index $n\geq 2$ because of its causal generation \cite{Durrer:2003ja}
\be
\vev{B_i(\bk)B_j^*(\bq)}=(2\pi)^3\de(\bk-\bq) (\de_{ij}-\hat{k}_i\hat{k}_j) \, A \, k^n~,
\ee
where $A$ is the amplitude of the spectrum. The magnetic energy momentum tensor is therefore automatically gauge invariant. Moreover, we work under the one-fluid MHD approximation, meaning that the conductivity of the universe is high, so that we can neglect the electric field and charge separation phenomena, occurring at very small scales. Since we are ultimately interested in the CMB spectra at large scales, we also neglect the presence of baryons and work under the tight coupling approximation. The universe is therefore composed only of a radiation component (photons and massless neutrinos - we also neglect neutrino masses), a pressure-less matter component (cold dark matter), and the magnetic field. 

In the first part of the paper, we solve Einstein's equations in the long wavelength limit, \ie consistently neglecting all terms proportional to $(k/\mathcal{H})^2$. By doing so, we can find analytic solutions for the gauge invariant variables describing the scalar metric perturbations and those of the total (radiation plus matter) fluid, which hold at scales larger than the horizon. Indeed, to calculate the Sachs Wolfe effect, we need the metric perturbations at scales larger than the horizon at recombination. 

The solutions are completely determined once the initial conditions have been specified. To set the initial conditions, we follow what has been done for the topological defect case \cite{Deruelle:1997py,Uzan:1998rt}. First of all, we assume a sudden phase transition, as for example the electroweak phase transition. The magnetic field is then instantaneously generated, if one considers perturbations with length scale relevant for the CMB today: $k / \mathcal{H}_{\rm B} \ll 1$, where $ \mathcal{H}_{\rm B}$ is the Hubble time during the phase transition, corresponding approximatively to its duration.
Before the phase transition, the universe is the usual perturbed FLRW universe with only the standard inflationary adiabatic perturbations; after the phase transition, the magnetic field energy momentum tensor contributes at first order to the metric perturbations. In order to connect these two stages, we match the geometry and the fluid variables on the surface of constant density, so that the induced three metric and the extrinsic curvature are continuous \cite{Deruelle:1997py}. This implies the conservation of the total energy momentum tensor at the magnetic field generation time. 

With this procedure, we find the solutions for the metric and fluid perturbation variables. Each variable is given by the sum of the adiabatic inflationary mode plus the magnetic field contribution, which is null before the transition. We have no freedom on how to add the magnetic field contribution: this is entirely specified by the matching. The solutions are such that the total fluid energy density and momentum are conserved at large scales, and the magnetic field does not alter the total curvature perturbation at large scales: this is given only by the inflationary contribution. Therefore, the matching procedure selects an isocurvature magnetic mode which leaves the curvature unchanged.

After neutrino decoupling, the solutions change again due to the non-zero neutrino anisotropic stress. In order to include analytically the free-streaming neutrinos in our analysis, we solve the Bardeen equation combined with the neutrino conservation equations, and find an analytical fit for the time evolution of the neutrino anisotropic stress. Then we introduce this fit back into Einstein's equation and solve for the metric and fluid perturbations including the contribution from the free-streaming neutrinos. With these complete, analytical solutions we then evaluate the Sachs Wolfe effect induced by the presence of the causal magnetic field. 

We confirm that the free-streaming neutrinos have a fundamental impact on the magnetic Sachs Wolfe \cite{Finelli:2008xh,PFP,Shaw:2009nf,yamafin}. Neglecting their presence one would conclude that the temperature anisotropy due to the magnetic field is proportional to $\Pi_\B/(k\eta_1)^2$, where $\eta_1$ denotes approximatively the conformal time at recombination. Therefore, the magnetic anisotropic stress would completely dominate the large scale Sachs Wolfe effect with respect to the magnetic energy density, and would induce a CMB spectrum of the type $\ell(\ell+1)\,C^\B_\ell\propto \ell^{-1}$. However, as already pointed out in \cite{Shaw:2009nf,Kojima:2009gw}, the neutrino anisotropic stress acts to compensate and reduce the magnetic field one. As a result, the Sachs Wolfe contribution correctly accounting for neutrinos becomes of the form $f(\eta_\B,\eta_\nu,\eta_{\rm rec})\,\Pi_\B/(k\eta_1)^2$, where $f(\eta_\B,\eta_\nu,\eta_{\rm rec})$ is a function of conformal time at recombination $\eta_{\rm rec}$, of the magnetic field generation time $\eta_\B$ and of the neutrino decoupling time $\eta_\nu$, which strongly suppresses the magnetic Sachs Wolfe contribution. Therefore, the contribution to the magnetic Sachs Wolfe behaving as $1/(k\eta_1)^2$ is only due to the evolution of the metric perturbations between the magnetic generation time and neutrino decoupling. 

This contribution has been neglected in the literature \cite{Finelli:2008xh,PFP,Shaw:2009nf,yamafin}, because it has been interpreted as connected to a decaying mode. However, with our analytical approach, we can show that the magnetic field solution does not behave as the standard inflationary one: one cannot really distinguish a growing and decaying mode, since all modes have a comparable amplitude at horizon crossing, and would therefore leave the same imprint on the CMB. The reason why this contribution is negligible only resides in the action of free-streaming neutrinos, the anisotropic stress of which counteracts the magnetic field one and effectively cancels the large scale temperature anisotropy. The new contribution to the Sachs Wolfe effect can only act as a source of anisotropy after the magnetic field generation time but before neutrino decoupling, and is negligible. Therefore, we finally confirm the result present in the literature \cite{Finelli:2008xh,PFP,Shaw:2009nf,yamafin}: the dominant contribution to the magnetic Sachs Wolfe effect becomes the one coming from the metric perturbations at the following order in the long wavelength expansion, \ie at order $(k\eta_1)^0$. This is what has been calculated in all the numerical analyses of the CMB temperature anisotropy. In the second part of the paper, we therefore proceed to evaluate analytically this contribution. 

In order to solve Einstein's equation at next-to-leading order, \ie including terms of the order $(k/\mathcal{H})^2$, the easiest way is to solve for the curvature perturbation equation. This allows us to find the solution for the Bardeen potentials at order $(k\eta_1)^0$, and consequently all the other fluid variables. We can then evaluate the Sachs Wolfe contribution at this order. 
 
The structure of the paper is the following: in section \ref{sec:Einstein} we derive Einstein's and conservation equations in the presence of a magnetic field. In section \ref{init_cond}, we explain the matching procedure, used then in section \ref{long} to calculate the metric and fluid variables at leading order in the $k\eta_1\ll 1$ expansion, both before and after neutrino decoupling. In section \ref{sec:SW} we compute the Sachs Wolfe effect from the leading order solution, and we conclude that the relevant contribution comes from the next-to-leading order. Therefore, in section \ref{nexttoleading} we calculate the next-to-leading order solutions, and we conclude in section \ref{sec:conc}. We consider scalar metric perturbations on a spatially flat Friedmann background, and we work with gauge invariant variables using the notations of \cite{KS} (with respect to \cite{Ma:1995ey}, we have $\Phi$ with the opposite sign while $\Psi$ is the same). We normalise the scale factor to one at equality, so that in a matter plus radiation universe $a=y^2+2y$ in terms of the dimensionless variable $y=\eta/\eta_1$, where $\eta$ denotes conformal time, $\eta_1=\eta_{\rm eq}/(\sqrt{2}-1)\simeq \eta_{\rm rec}$ and $\eta_{\rm eq}$ represents conformal time at equality. Scales larger than the horizon at recombination satisfy $x_1 \equiv k\eta_1\ll 1$. A dot denotes derivative with respect to conformal time, while a prime denotes derivative with respect to $y$. Greek indexes go from 0 to 3, while latin ones from 1 to 3. The neutrino background energy density fraction is denoted $R_\nu=\bar\rho_\nu/\bar\rho_{\rm rad}\simeq 0.4$.

\section{Einstein's and conservation equations}
\label{sec:Einstein}

In this section we derive Einstein's and conservation equations with a non-zero magnetic field. We assume that the electromagnetic energy momentum tensor is first order in perturbation theory, meaning that the fields themselves are half order. The electromagnetic field tensor and the energy momentum tensor can therefore be defined with respect to the unperturbed velocity of the fluid energy frame of the FLRW background, $\bar{u}^\al $
\bea
E^\al &=& F^{\al\b}\bar{u}_\b~, \\
B^\al &=& \frac{1}{2} {\epsilon^\al}_{\b\ga}F^{\b\ga}~, \\
{T_{\rm em}^\al}_\b &=& (\rho_{\rm em}+p_{\rm em}) \, \bar{u}^\al \, \bar{u}_\b+ p_{\rm em} \, {{\bar g}^\al}_{\;\; \b}+ 2\, \bar{u}^{(\al} {q^{\rm em }}_{\b)}+{\pi_{\rm em}^\al}_\b~, 
\eea
where $\epsilon_{\al\b\ga} = \de_{[\al}^1 \de_\beta^2 \de_{\gamma]}^3 $ is the totally antisymmetric rank 3 tensor. The components of the energy momentum tensor are
\bea
\rho_{\rm em} &=& \frac{1}{2}(E^2+B^2)~, \\
p_{\rm em} &=& \frac{1}{6}(E^2+B^2)~, \\
q_{\rm em}^\al &=& {\ep^\al}_{\b\ga}E^\b B^\ga~,  \\
{\pi_{\rm em}^\al}_{\b} &=& \frac{E^2+B^2}{3} {\bar{g}^\al}_{\;\;\b} + \frac{E^2+B^2}{3} \bar{u}^\al\bar{u}_\b-E^\al E_\b-B^\al B_\b~. 
\eea 
From Maxwell's equations ${F^{\al\b}}_{;\b}=j^\al$ one sees that the current density $j^\al$ is also half order as the electromagnetic field. The covariant Ohm's law can therefore be written with respect to the background velocity
\be
j^\al+\bar u^\al \bar u_\b j^\b=\sigma F^{\al\b} \bar u_\b~.
\ee 
We work under the MHD approximation, \ie the conductivity $\sigma$ is infinite. From the above equation one sees that, in order to keep the current finite, the electric field must vanish \cite{Jackson}. Therefore in the following we set the electric field to zero. The MHD approximation is valid on sufficiently large scales, where charge separation effects are not important. 

The total energy momentum tensor includes the fluid, labelled by ${}_{\rm F}$ and representing radiation and matter, and the magnetic field. Using the notation $\rho_{\rm B}=B^2/2$, $p_{\rm B}=B^2/6$, ${\pi_{\rm B}^i}_j=(B^2/3) \bar g^i_j - B^i B_j$, the components of the total energy momentum tensor  in real space are
\bea
{T^0}_0 &=&-\bar\rho_{\rm F} -\delta \rho_{ \rm F}-\rho_{\rm B}~, \\
{T^i}_0 &=& - a (\bar\rho_{ \rm F}+\bar p_{ \rm F}) \de u^i~, \\
{T^i}_j &=& (\bar p_{\rm F}+\de p_{\rm F}+p_{\rm B})  \de^i_j + \bar p_{\rm F} \, {{\pi_{\rm F}}^i}_j + {{\pi_{\rm B}}^i}_j~.
\eea
We only consider scalar perturbations. Following \cite{KS}, we expand the scalar part of the metric and fluid perturbations by scalar harmonic functions $Y=e^{-{\rm i}\bk\cdot \bx}$ (for the definition of $Y^i$ and ${Y^i}_j$ see \cite{KS}). The perturbed scalar energy momentum tensor in wave-number space becomes then
\bea
\de {T^0}_0 &=& (-\delta \rho_{ \rm F}-\rho_{\rm B}) Y~,  \\
\de {T^i}_0 &=& - (\bar\rho_{ \rm F}+\bar p_{ \rm F}) \, v \, Y^i~, \\
\de {T^i}_j &=& (\de p_{\rm F}+p_{\rm B})  \de^i_j + ( \bar p_{\rm F} \, \pi_F + \pi_{\rm B}) {Y^i}_j~,
\eea
where $v$ is the scalar part of the velocity perturbation $\de u^i$ and (see~\cite{Kahniashvili:2006hy,Brown:2005kr})
\bea
\rho_{\rm B} &=& \frac{1}{2}  (B_i*B^i)~, \\
\pi_{\rm B} &=& \frac{3}{2}\hat{k}^i\hat{k}_j( B_i * B^j ) - \frac{1}{2} (B_m*B^m)~,  \\
B_i*B^j &=& \int \frac{d^3 p}{(2\pi)^3} B_i(\bp)B^j(\bk-\bp)~.
\eea
The scalar perturbed Einstein equations $\delta {G^\mu}_\nu=8\pi G \delta {T^\mu}_\nu$ in terms of gauge invariant variables, in the presence of the primordial magnetic field and in a spatially flat universe \cite{Kahniashvili:2006hy,KS} are
\bea
k^2\Phi &=& 4\pi G a^2 \bar\rho_{\rm F} (D+\frac{\rho_{\rm B}}{\bar\rho_{\rm F}})~, \label{Einstein1} \\
-k^2(\Phi+\Psi) &=& 8\pi G a^2\bar{p}_{\rm F} (\pi_{\rm F}+\frac{\pi_{\rm B}}{\bar{p}_{\rm F}})~, \label{Einstein2} \\
k(\mathcal{H}\Psi-\dot\Phi) &=& 4\pi G a^2 (\bar\rho_{\rm F}+\bar p_{\rm F}) \,V~, \label{Einstein3}
\eea
where $\Phi$ and $\Psi$ are the Bardeen potentials, $D$ is the gauge invariant variable corresponding to the density perturbation in the velocity-orthogonal slicing, and $V$ is the gauge invariant variable corresponding to the velocity perturbation in the Newtonian longitudinal gauge \cite{KS}. 

The conservation equations ${T_{\rm F}^\al}_{\b \, ; \al} + {T_{\rm B}^\al}_{\b \, ; \al}=0$ involve the momentum exchange $Q_i$ between the fluid and the magnetic field, represented by the Lorentz force (the energy exchange is set to zero since it involves the electric field). Projecting the conservation equation on the rest space of the comoving observer by means of the projector ${h^\al}_\b={{\bar g}^\al}_{\;\;\b}+\bar u^\al \bar u_\b$ gives
\be
{h_{\al}}^\b {T_{\rm F}}^\ga_{\b\, ; \ga}=-{h_{\al}}^\b {T_{\rm B}}^\ga_{\b \, ; \ga}=\epsilon_{\al \ga \de} J^\ga B^\de \equiv Q_\al~, 
\label{conserv}
\ee 
where $Q_\alpha$ is a purely spatial vector ($Q_0=0$) denoting the Lorentz force. We Fourier transform it and extract its scalar part by 
$-{\rm i} \, \hat{k}^i Q_i= \ell_{\rm B}\,Y$, which from \eqref{conserv} must satisfy the relation
\bea
\frac{\rho_{\rm B}}{2}& =& \pi_{\rm B}+\frac{3}{2}\frac{\ell_{\rm B}}{k}~,  \label{sum} \\
\ell_{\rm B} &=& \frac{k}{2} (B_i * B^i)- k^j \hat{k}_l (B_j *  B^l)~.
\eea
The fluid conservation equations become
\bea
\dot D-3w \HH D &=& -k(1+w)V-2w \HH \pi_{\rm F} +\frac{3\HH}{k \bar \rho_{\rm F}} \ell_{\rm B}~,  \label{consen} \\
\dot V +\HH V &=& k\Psi +\frac{k \, c_s^2}{1+w} D -\frac{2}{3} \frac{w}{1+w} k \, \pi_{\rm F}+\frac{\ell_{\rm B}}{\bar \rho_{\rm F}(1+w)}~, \label{consmom}
\eea
where $w=\bar p_{\rm F}/\bar \rho_{\rm F}$, $c_s^2=\dot{\bar p}_{\rm F}/\dot{\bar \rho}_{\rm F}$ and the fluid has no internal entropy perturbation $\Gamma_{\rm F}=0$ (we remind that we have set the energy exchange between the fluid and the magnetic field to zero). We introduce the following notations
\be
\Om_{\rm B} = \frac{\rho_{\rm B}}{\bar \rho_{\rm rad}} ~~~~~
\Pi_{\rm B} = \frac{\pi_{\rm B}}{\bar \rho_{\rm rad}} ~~~~~
L_{\rm B} = \frac{\ell_{\rm B}}{\bar \rho_{\rm rad}} ~,
\ee
with $\bar \rho_{\rm F}=\bar \rho_{\rm rad} + \bar \rho_{\rm mat}$, so that \eqref{sum} becomes $\Om_{\rm B}=2\Pi_{\rm B}+3 L_{\rm B}/k$. 

\section{Matching conditions}
\label{init_cond}

We assume that the magnetic field is generated by a causal process acting `fast', \ie within one Hubble time, as for example a sudden phase transition in the early universe. We denote the magnetic field generation time by $\eta_{\rm B}$. Before this time, only adiabatic perturbations of inflationary origin are present in the universe. Afterwards, there is an extra contribution to the metric perturbations due to the magnetic field, which in turns affects the fluid perturbations. In order to establish the initial conditions for this system, we follow what has been done in \cite{Deruelle:1997py} for the analogous case of the topological defects. They perform a matching of the perturbations on a constant energy density surface, which allows to make the link between the pre- and post-magnetic field phases in the correct way: the physical configuration satisfies the conservation of the total energy momentum tensor (and consequently Einstein's equations) at all times. The two independent matching conditions are given by Eqs.~(36) of \cite{Deruelle:1997py}, and read 
\bea
\left[ 8 \pi G \, a_{\rm B}^2 \, (\bar \rho_{\rm F} D_g +\rho_{\rm B} )\right]_\pm &=& 0~, \\
\left[ \Phi \right]_\pm &=& 0~,
\eea
where
\be
F_\pm= \lim_{\epsilon\to 0} \, [F(\eta_{\rm B}+\epsilon)-F(\eta_{\rm B}-\epsilon)]~,
\ee
and 
\be
D_g=D-3(1+w)\frac{\HH}{k}V+3(1+w)\Phi~.
\ee
In terms of the variable $D$ and using \eqref{Einstein1} we obtain then the conditions
\bea
D(\eta_\B+\ep) + \frac{\Om_\B}{1+a_\B} & =& D(\eta_\B-\ep)~, \label{matchD}\\
V(\eta_\B +\ep) &=& V(\eta_\B-\ep)~. \label{matchV}
\eea
We now proceed to solve the system of Einstein's plus conservation equations at large scales. We need to split the universe evolution in three stages:  before magnetic field generation, after magnetic field generation and after neutrino decoupling. For the transition from the first to the second stage we impose the matching conditions derived here, while neutrino decoupling does not introduce any discontinuity in the metric and fluid perturbation variables.

\section{Leading order solutions at large scales $x_1\ll 1$}
\label{long}

In order to solve the system of Einstein's plus conservation equations, we choose to derive a second order differential equation for the variable $D$. We combine the total density and momentum conservation equations (\ref{consen}) and (\ref{consmom}) to get
\bea
 && \ddot{D}+(1+3c_s^2 -6w)\HH \dot D +3 \HH^2 \left[ -\frac{1}{2}-4w+\frac{3}{2}w^2 +3c_s^2 +\frac{c_s^2}{3} \left( \frac{k}{\HH}\right)^2 \right] D = \nonumber \\
& & 2 \HH ^2 \left[ -2w+3c_s^2+3w^2+\frac{w}{3}  \left( \frac{k}{\HH}\right)^2 \right] \pi_{\rm F} -2\HH w \dot\pi_{\rm F} + \nonumber \\
& & \frac{\HH^2}{1+a} \left[ 2-3w+3c_s^2-\frac{a}{1+a} -\frac{1}{3} \left( \frac{k}{\HH}\right)^2 \right] \Om_\B +\nonumber \\
& &\frac{2 \HH^2}{1+a} \left[ 1+6w-3c_s^2+\frac{a}{1+a}+\frac{1}{3} \left( \frac{k}{\HH}\right)^2 \right] \Pi_\B~.
\label{Dddot}
\eea
In order to determine the Sachs Wolfe effect, we only need to solve for scales which are over the horizon at recombination, $x_1\equiv k\eta_1\ll 1$. Therefore, in a first instance, we drop the terms proportional to $(k/\HH)^2$.

\subsection{Before neutrino decoupling}
\label{without_n}

Before their decoupling at a temperature of $T_{\nu}\simeq 1$ MeV, neutrinos do not free-stream and have therefore zero anisotropic stress. They are characterized only by their background energy density $\bar{\rho}_\nu$, their density perturbation $D_{\nu}$ and their velocity perturbation $V_{\nu}$. These are simply included in the total matter perturbations $D$ and $V$. Consequently, for $T>T_{\nu}$ the fluid anisotropic stress $\pi_{\rm F}$ in Eq.~(\ref{Dddot}) vanishes. 

In order to solve Eq.~(\ref{Dddot}) in a matter plus radiation universe with $\bar \rho_{\rm F}=\bar \rho_{\rm rad} + \bar \rho_{\rm mat}$, we follow \cite{libroruth}. Before magnetic field generation the source in Eq.~(\ref{Dddot}) drops, and we have the usual homogeneous solution, that we express here as a function of the variable $y=\eta/\eta_1$
\bea
D^{-}(y)&=&a_1 u_{\rm R} (y) + a_2 u_{\rm S}(y)~,\label{Dmoins} \\
u_{\rm S}(y)&=& \frac{1}{2y+3y^2+y^3}~, \\
u_{\rm R}(y)&=& u_{\rm S}(y) \left[ \frac{y^3(8+3y)(10+10y+3y^2)}{9(1+y)} \right]~,
\eea
where $a_1$ and $a_2$ are arbitrary constants and we set the decaying mode to zero: $a_2=0$. We remind here that in terms of $y$ the scale factor is $a=y^2+2y$ and therefore the Hubble parameter is $\HH^2= 4(1+a)/(\eta_1^2 a^2)=4(1+y)^2/[\eta_1^2(y^2+2y)^2]$, the equation of state parameter of the total fluid is $w(y)=1/[3(1+y)^2]$, and the total sound speed is $c_s^2(y)=4/[3(4 + 6 y + 3 y^2)]$. 

In order to solve after the generation of the magnetic field we use the Wronskian method, where the source is given by the right hand side of \eqref{Dddot} with $\pi_{\rm F}=0$ in terms of $y$
\be
S_\B(y)=\frac{(1+y)^4\Big[(1+y)^2(8+6y+3y^2)\Omega_\B+ 2(8+26y+37y^2+24y^3+6y^4)\Pi_\B \Big]}{y^2(2+y)^2(4+6y+3y^2)}~.
\label{SB}
\ee 
The solution becomes then
\bea
D^{+}(y)&=& [b_1 + B_1 (y) ] u_{\rm R} (y) + [b_2 + B_2 (y)] u_{\rm S}(y)~, ~~~~~~~{\rm where} \label{Dplus} \\
W(y)&=& u_{\rm R} \frac{du_{\rm S}}{dy} - u_{\rm S} \frac{du_{\rm R}}{dy}~, \\
B_1(y)&=& -\int dy \, \frac{u_{\rm S} S_\B}{ W}~, \\
B_2(y)&=& \int dy \,\frac{u_{\rm R} S_\B}{W}~,
\eea
where again $b_1$ and $b_2$ are arbitrary constants. The metric perturbations $\Phi$ and $\Psi$, and the velocity perturbation $V$ can now be calculated using Einstein's equations (\ref{Einstein1}) to (\ref{Einstein3}) both before the magnetic field generation, setting $\rho_\B=\pi_\B=0$ and using $D^{-}$ given in Eq.~(\ref{Dmoins}), and after the magnetic field generation, using instead Eq.~(\ref{Dplus}). Consequently, $\Phi$, $\Psi$ and $V$ after the magnetic field generation are functions of $b_1$ and $b_2$. We then use the matching conditions \eqref{matchD} and (\ref{matchV}) to determine the free parameters $b_1$ and $b_2$. This completely specifies the magnetic contribution to the metric and fluid perturbations without ambiguity. The only free parameter remains $a_1$, the amplitude of the scalar metric perturbation from inflation, which is then determined by the COBE normalisation. Denoting $y_\B$ the magnetic field generation time, the matching gives
\be
b_1=a_1-\frac{3}{5}\Om_\B~~~~~{\rm and}~~~~~b_2=\frac{16}{15}\Om_\B+4(1+y_\B)\Pi_\B~,
\ee
so that
\be
D^+(y)=\frac{y^2(8+3y)(10+10y+3y^2)}{9(2+y)(1+y)^2}\,a_1 - \frac{\Om_\B}{(1+y)^2} - \frac{4(y-y_\B)}{y(2+y)(1+y)}\Pi_\B~.
\label{Dplusy}
\ee
The first part of the above expression, proportional to $a_1$, is the inflationary mode of \eqref{Dmoins}, while the second part proportional to $\Om_\B$ and $\Pi_\B$ is the magnetic field contribution.  For later convenience, the magnetic field contribution is denoted $D^\B$, so that $D^+(y)=D^-(y)+D^\B(y)$. Note that $D^\B(y)$ is a decaying mode, but its amplitude is of order $\Om_\B+2\Pi_\B$ as long as $y\lesssim 1$, {\it i.e.} anytime before recombination (in the conclusion section \ref{sec:conc}, we discuss the role of the decaying mode of the perturbations generated by the magnetic field). The metric perturbations $\Phi$ and $\Psi$, and the velocity $V$ follow from (\ref{Einstein1}-\ref{Einstein3}), and are given in appendix \ref{app1}. 

Note that these solutions hold at large scales $k\ll 1/\eta_1$. This is because in \eqref{Dddot} we have dropped the terms proportional to $(k/\HH)^2$ with respect to those constant in $k$. A consequence of this is, in particular, that the momentum conservation equation (\ref{consmom}) is only satisfied at leading order in $x_1=k\eta_1$: \ie neglecting the Lorentz force which is order $\mathcal{O}(x_1^0)$ and the term proportional to $k \,c_s^2\, D$ which is order $\mathcal{O}(x_1)$, and keeping only the term $k \, \Psi$ which is order $\mathcal{O}(1/x_1)$.

\subsection{After neutrino decoupling}
\label{with_n}

After decoupling, neutrinos start to free-stream, and acquire a non-zero anisotropic stress $\pi_\nu$. Consequently, for $\eta > \eta_\nu$ Eq.~(\ref{Dplusy}) gets a new source term, generated by $\pi_{\rm F}=R_\nu\pi_\nu$ (where $R_\nu\equiv \bar{\rho}_\nu/\bar{\rho}_{\rm rad}$). Since, however, the neutrinos do not add as a new component in the universe (as did the magnetic field), in this case we do not need to perform a matching to guarantee the conservation of the total energy momentum tensor. The new anisotropic stress component does not introduce a discontinuity, since it builds up at sub-horizon scales continuously in time, so that $\pi_\nu(y_\nu)=0$. We have then
\be
D(y)=\left\{ \begin{array}{ll} D^-(y)&\quad \hbox{for}\quad y<y_B~,\\
 D^+(y)=D^-(y)+D^\B(y) &\quad  \hbox{for} \quad y_B<y<y_\nu~, \\
 D^{\rm fin}(y)\equiv D^+(y) +\Dst(y)  &\quad  \hbox{for} \quad y>y_\nu~, \\
\end{array} \right.
\label{Dfin}
\ee
where $D^-(y)$ is the inflationary solution Eq.~(\ref{Dmoins}) with $a_2=0$, $D^+(y)$ is given in Eq.~(\ref{Dplusy}), and $\Dst(y)$ is generated by the neutrino anisotropic stress, such that $\Dst(y_\nu)=0$. $\Dst(y)$ is calculated again using the Wronskian method, where now the source is given by the fluid anisotropic stress in Eq.~(\ref{Dddot})
\be
S_\nu(y)=-2(\HH\eta_1)^2\Big[-2 w+3c_s^2+3w^2\Big]R_\nu\pi_\nu(y)
+2\HH\eta_1 wR_\nu\frac{d\pi_\nu}{dy}~.
\ee
We again consistently neglect the terms proportional to $(k/\HH)^2$. The solution becomes then
\bea
\Dst(y)&=&  C_1 (y)  u_{\rm R} (y) +  C_2 (y) u_{\rm S}(y)~, ~~~~~~~{\rm where} \label{Dst} \\
C_1(y)&=& -\int_{y_\nu}^y dy' \, \frac{u_{\rm S}(y') S_\nu(y')}{ W(y')}  \quad \mbox{and} \quad C_2(y)= \int_{y_\nu}^{y} dy' \,\frac{u_{\rm R}(y') S_\nu(y')}{W(y')}~.  \label{C1C2}
\eea
As explained above, since the source $S_\nu(y)$ is continuous, the energy-momentum conservation is automatically satisfied at the neutrino decoupling time $y_\nu$, and no additional matching conditions are needed. Hence once the neutrino anisotropic stress $\pi_\nu(y)$ is determined, the solution $D^{\rm fin}$ is completely fixed.

The evolution of $\pi_\nu$ in the presence of an external constant anisotropic stress has been studied in \cite{Kojima:2009gw}  (see also \cite{Shaw:2009nf}). Even though the neutrinos and the magnetic field do not interact directly but only through gravity, the neutrino anisotropic stress quickly adjusts to the external one and compensates it. In order to determine the time evolution of $\pi_\nu$, we follow the method given in \cite{Shaw:2009nf}, \ie we combine the neutrino conservation equations and Einstein's equations to derive a fourth order differential equation for $\pi_\nu(y)$. This derivation is presented in appendix~\ref{app:pinu}. The final time dependence of $\pi_\nu(y)$ is rather complicated, but it can be approximated well by (\cf Fig.~\ref{fig:Pinu}) 
\be
\label{pinuapp}
\pi_\nu(y)=\frac{3\Pi_\B}{R_\nu}\left(\frac{y_\nu^2}{y^2}-1 \right)-\frac{40 \, a_1}{15+4R_\nu}y(y-y_\nu)~.
\ee
\FIGURE{\epsfig{file=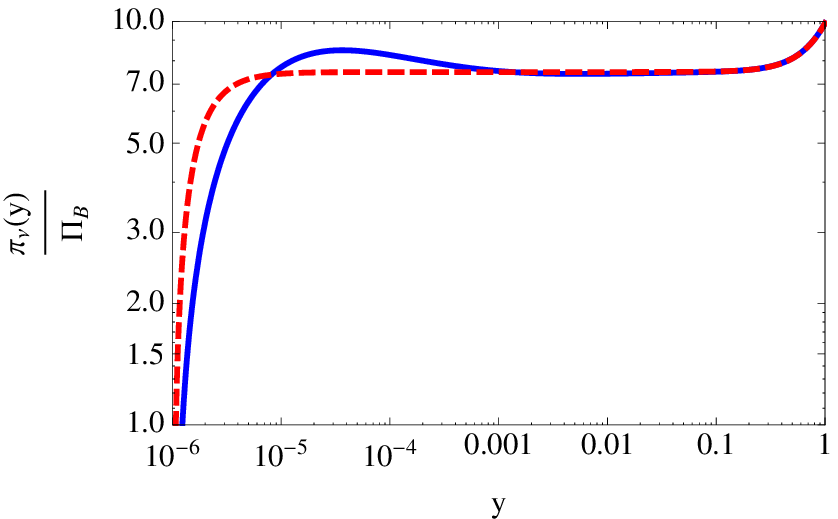,width=10cm} 
        \caption{Blue, solid line: the neutrino anisotropic stress as a function of $y$ derived in appendix \ref{app:pinu}, Eq.~(\ref{pinusol}), and red, dashed line: the fit given in \eqref{pinuapp}. We have chosen comparable values for $\Pi_\B\simeq a_1$, and normalised by $\Pi_\B$. The two curves flatten at $\pi_\nu(y)\rightarrow -3\Pi_\B/R_\nu$. The growth at $y\simeq 1$ represents the inflationary contribution to the neutrino anisotropic stress, \ie the second term in the right hand side of \eqref{pinuapp}. The time at which this contribution starts to dominate depends on the relative amplitude of $\Pi_\B$ and $a_1$. }
	\label{fig:Pinu}}
With this approximation, we can solve analytically the integrals in \eqref{C1C2}. We find then
\be
\label{Dstress}
\Dst(y)=\frac{80(2 y + y_\nu)(y - y_\nu)^2}{9(15+4 R_\nu) y (1+y)(2+y)}R_\nu \,a_1+\frac{4 (y - y_\nu)^2} {y^2 (1+y)(2+y)}\, \Pi_\B~.
\ee
The part proportional to $a_1$ is the standard neutrino contribution to the inflationary perturbation: we denote it $\Dst_{\rm inf}$. The part proportional to $\Pi_\B$ is the contribution due to the interplay among the magnetic and neutrino anisotropic stresses: we denote it $\Dst_{\rm mag}$, so that $\Dst(y)=\Dst_{\rm inf}(y)+\Dst_{\rm mag}(y)$. The metric perturbations $\Phist$ and $\Psist$, and the velocity $\Vst$ are again given in appendix \ref{app1}. It is clear that the time dependence of \eqref{Dstress}, and consequently of the other metric and fluid perturbations given in the appendix, follows directly from our choice for the approximated form of $\pi_\nu(y)$ given in \eqref{pinuapp}. Since this fit interpolates well the behaviour of the real neutrino anisotropic stress, these expressions will give the right order of magnitude for the evaluation of the Sachs Wolfe effect. However, their detailed time dependence is to be considered only indicative. 

Summarising, we have that the final metric and fluid perturbations valid after neutrino decoupling can be decomposed into three contributions: the standard inflationary one, the magnetic one arising after magnetic field generation, and the one due to the free-streaming neutrinos which further contribute after their decoupling to the metric perturbations through their anisotropic stress
\bea
D^{\rm fin}(y>y_\nu)&=& D^-(y)+D^\B(y)+\Dst(y)~, \\
\Phi^{\rm fin}(y>y_\nu)&=& \Phi^-(y)+\Phi^\B(y)+\Phist(y)~, \\
\Psi^{\rm fin}(y>y_\nu)&=& \Psi^-(y)+\Psi^\B(y)+\Psist(y)~, \\
V^{\rm fin}(y>y_\nu)&=& V^-(y)+V^\B(y)+\Vst(y)~. 
\eea

\vspace*{1cm}

\noindent
Using the above solutions we can calculate the curvature perturbation $\zeta$. By imposing the continuity of the induced three metric and the extrinsic curvature on the surface of constant density, the matching conditions guarantee in particular that $\zeta$ is continuous at the magnetic field generation time $\eta_\B$. Consequently, the magnetic contribution to $\zeta$ is by definition such that it vanishes at $\eta_\B$, and only the constant inflationary part remains. Moreover, if we calculate $\zeta$ for $\eta>\eta_\B$ using the solutions for $\Phi$ and $\Psi$ derived above, we further find that the magnetic contribution to it vanishes at all times
\be
\zeta(y>y_\B)= -\Phi^{+} +\frac{2}{3(1+w)} \left( \Psi^+ -\frac{\dot{\Phi}^+}{\HH} \right)  =-\frac{10\,a_1}{x_1^2}~~~~\Leftrightarrow~~~~\frac{d\zeta}{dy}=0~.
\label{zeta}
\ee
This shows that the curvature perturbation is conserved at large scales $x_1\ll 1$: therefore, if the magnetic contribution is zero at $\eta_\B$, it must remain so for $\eta>\eta_\B$ at leading order in $x_1$, \ie $\zeta\propto\mathcal{O}(1/x_1^2)$. The neutrinos also do not contribute to the curvature perturbation, since we have imposed adiabatic initial conditions for the inflationary solution 
\be
\zeta(y>y_\nu)= -\Phi^{\rm fin} +\frac{2}{3(1+w)} \left( \Psi^{\rm fin} -\frac{\dot{\Phi}^{\rm fin}}{\HH} \right)  =-\frac{10\,a_1}{x_1^2}~.
\label{zetanu}
\ee
We will see in the following that the curvature is no longer conserved for intermediate scales, \ie at the following order in the $x_1\ll 1$ expansion, $\zeta\propto\mathcal{O}(x_1^0)$ (\cf section \ref{nexttoleading}).

\section{Sachs Wolfe effect from the leading order solution}
\label{sec:SW}

In the approximation of instantaneous recombination, valid for wavelengths larger than the Hubble scale at recombination $x_1\ll 1$, the Sachs Wolfe contribution to the temperature anisotropy is (see \eg \cite{libroruth})
\be
\frac{\De T}{T}(k, \eta_0) \simeq \frac{D_{g\,\gamma}(k, \eta_{\rm rec})}{4} +\Psi(k, \eta_{\rm rec})-\Phi(k, \eta_{\rm rec})~.
\label{SW}
\ee
In order to evaluate $D_{g\,\gamma}$, knowing the Bardeen potentials $\Phi$ and $\Psi$ which have been derived above, we use the conservation equations for the photon fluid
\bea
{\dot{D}}_{g\,\gamma} &=& -\frac{4}{3} k V_{\gamma}~, \label{consenrad} \\
{\dot{V}}_{\gamma}  &=& k (\Psi-\Phi) +\frac{k }{4} D_{g\,\gamma} +\frac{3}{4(1-R_\nu)} L_{\rm B}~. \label{consmomrad}
\eea
These follow directly from (\ref{consen}) and (\ref{consmom}). We remind that we work under the tight coupling approximation, so that we neglect the Thomson scattering term, we equal the baryons and photon velocities $V_\gamma\simeq V_{\rm b}$, and neglect the baryons energy density so that the parameter $R_{\rm b}=3\rho_{\rm b}/4\rho_\gamma\equiv 0$. Note that in tight coupling the momentum exchange due to the Lorentz force must be included in the photon momentum conservation equation, since it acts on baryons and $V_\gamma\simeq V_{\rm b}$. From the above equations it follows that the Sachs Wolfe contribution is 
\be
\frac{\De T}{T}\simeq\frac{D_{g\,\gamma}(y_{\rm rec})}{4} +\Psi(y_{\rm rec})-\Phi(y_{\rm rec}) = \frac{1}{k} \left [\dot{V}_{\gamma} -\frac{3}{4(1-R_\nu)}  L_\B  \right]_{\rm rec}~;
\label{SWV}
\ee
we therefore only need to evaluate $V_{\gamma}$. In order to do this, we differentiate \eqref{consmomrad} and substitute with \eqref{consenrad} (note that $L_\B=k(\Om_\B-2\Pi_\B)/3$ is constant) to obtain a second order differential equation for $V_{\gamma}$, which reads in terms of the variable $y$
\be
\frac{d^2 V_{\gamma}}{dy^2}+\frac{x_1^2}{3}V_{\gamma}=x_1 \frac{d}{dy}(\Psi-\Phi)~.
\label{DVgamma}
\ee
Following what found in the previous sections and presented in the end of section \ref{with_n}, it appears that the source term can be divided into the standard inflationary contribution, the magnetic contribution, and the free-streaming neutrino contribution (\cf Eqs.~(\ref{Phiplus}, \ref{Psiplus}) and (\ref{Phistress}, \ref{Psistress}) of appendix \ref{app1}). Consequently, $V_\gamma$ can also be divided as
\be
V_{\gamma} = V_{\gamma}^-+V_{\gamma}^\B+\Vst_{\gamma} ~.  \label{Vsum}
\ee
The standard inflationary contribution $V_\gamma^-$ can be determined by solving directly the differential equation (\ref{DVgamma}), where the right hand side is given by $\Psi^-(y)$ and $\Phi^-(y)$ (\cf Eqs.~(\ref{Phiplus}) and (\ref{Psiplus}) of appendix \ref{app1}). The magnetic and free-streaming neutrino contributions can be determined instead using the Wronskian method, and after one integration by parts of the source integrals they can be written as
\bea
V_{\gamma}^\B & = &  x_1 \cos\left (\frac{x_1 y}{\sqrt{3}}\right )  \int_{y_\B}^y dy' (\Psi^\B-\Phi^\B) \cos \left (\frac{x_1 y'}{\sqrt{3}}\right ) \nonumber \\
& & + \, x_1 \sin\left (\frac{x_1 y}{\sqrt{3}}\right ) \int_{y_\B}^y dy' (\Psi^\B-\Phi^\B) \sin \left (\frac{x_1 y'}{\sqrt{3}}\right )~, \label{VB} 
\eea

\bea
\Vst_{\gamma} & = & x_1  \cos\left (\frac{x_1 y}{\sqrt{3}}\right )  \int_{y_\nu}^y dy' (\Psist-\Phist) \cos \left (\frac{x_1 y'}{\sqrt{3}}\right )  \nonumber \\
& & + \, x_1 \sin\left (\frac{x_1 y}{\sqrt{3}}\right ) \int_{y_\nu}^y dy' (\Psist-\Phist) \sin \left (\frac{x_1 y'}{\sqrt{3}}\right )~. \label{Vgstress} 
\eea
Note that the boundary of the integrals in Eqs.~ (\ref{VB}) and (\ref{Vgstress}) are a consequence of the fact that $V_\gamma$ is continuous both at magnetic field generation time $y_\B$, as implied by the matching conditions, and at the neutrino free-streaming time $y_\nu$. 

\vspace*{1cm}

\noindent
As an example we first evaluate the standard inflationary contribution to the Sachs Wolfe effect, given by $V^-_\gamma$. We impose adiabatic initial conditions, such that at large scales $x_1\ll 1$ we have $V_{\gamma}(y) \simeq V_{\nu}(y)\simeq V_{\rm mat}(y) \simeq V^{-}(y) $, where $V^{-}$ denotes the total velocity perturbation from inflationary perturbations given in \eqref{Vplus}. The initial conditions completely specify the arbitrary constants $c_1$ and $c_2$ coming from the integration of \eqref{DVgamma}. To the standard inflationary contribution we should further add the free-streaming neutrino contribution, represented by the part of \eqref{Vgstress} sourced by $\Psist_{\rm inf}-\Phist_{\rm inf}$. This is usually neglected at large scales, since it has an impact only of a few percent on the standard inflationary Sachs Wolfe: we evaluate it in appendix \ref{app:neutrinosSW}. From \eqref{SWV}, the inflationary Sachs Wolfe effect is then given by
\be
\left. \frac{\De T}{T} \right|_{\rm inf} \simeq \frac{\dot{V}^{-}(y_{\rm rec})}{k}~.
\ee
Using \eqref{Vplus} we find
\be
\left. \frac{\De T}{T} \right|_{\rm inf} \simeq - \frac{2(40 + 40 y_{\rm rec} + 18 y_{\rm rec}^2 + 3 y_{\rm rec}^3)}{3 (2 + y_{\rm rec})^3}\,\frac{a_1}{x_1^2} \simeq -\frac{2.5\,a_1}{x_1^2}~. 
\label{dToverTinf}
\ee
Comparing with \eqref{Phiplus}, one sees that this result corresponds to the usual $\Phi^-/3$.  

\vspace*{1cm}

\noindent
We now calculate the Sachs Wolfe effect due to a non-zero primordial magnetic field. Combining Eqs.~(\ref{SWV}) and (\ref{Vsum}) this reads, in terms of the variable $y$ 
\be
\left. \frac{\De T}{T} \right|_\B \simeq  \frac{1}{x_1} \left [\frac{d V_\gamma^\B}{dy}+\frac{d \Vst_\gamma}{dy} \right]_{y_{\rm rec}} 
-\frac{3}{4(1-R_\nu)}  \frac{L_\B}{k}~, 
\ee
where for $\Vst_\gamma$ we only take into account the magnetic ($\Vst_\B$) and not the inflationary part ($\Vst_{\rm inf}$, calculated in appendix \ref{app:neutrinosSW}). Differentiating Eqs.~(\ref{VB}) and (\ref{Vgstress}) and further writing the Lorentz force as $L_\B/k=\Om_\B/3+2\Pi_\B/3$ we finally find 
\bea
\left. \frac{\De T}{T} \right|_\B &\simeq & (\Psi^\B-\Phi^\B)+(\Psist_{\rm mag}-\Phist_{\rm mag}) -\frac{1}{4(1-R_\nu)}(\Om_\B-2\Pi_\B) 
\label{dToverTBcomplete} \\
&+& \frac{x_1}{\sqrt{3}}  \cos\left (\frac{x_1 y}{\sqrt{3}}\right ) \left[  \int_{y_\B}^{y} dy' (\Psi^\B-\Phi^\B) \sin \left (\frac{x_1 y'}{\sqrt{3}}\right ) +\int_{y_\nu}^{y} dy' (\Psist_{\rm mag}-\Phist_{\rm mag}) \sin \left (\frac{x_1 y'}{\sqrt{3}}\right ) \right] \nonumber \\
&-&  \frac{x_1}{\sqrt{3}} \sin \left (\frac{x_1 y}{\sqrt{3}}\right ) \left[ \int_{y_\B}^{y} dy' (\Psi^\B-\Phi^\B) \cos \left (\frac{x_1 y'}{\sqrt{3}}\right )  
+ \int_{y_\nu}^{y} dy' (\Psist_{\rm mag}-\Phist_{\rm mag}) \cos \left (\frac{x_1 y'}{\sqrt{3}}\right )  \right]~. \nonumber
\eea
The leading term at large scales $x_1\ll 1$ comes from the first two terms of the above equation, which can be derived from Eqs.~ (\ref{Phiplus}) and (\ref{Psiplus}). We immediately see that it is of order $\mathcal{O}(1/x_1^2)$, while both the Lorentz force term and the integrals are of order $\mathcal{O}(x_1^0)$. Let us then concentrate on this apparently leading term. For the purely magnetic part, we find 
\be
\Psi^\B-\Phi^\B=\frac{12\, [3y^2+2y-4(1+y)y_\B] }{y^3(2+y)^3} \, \frac{\Pi_\B}{x_1^2}~.
\label{PsimoinsPhiB}
\ee
Supposing that the magnetic field is generated at the EW phase transition at $T\simeq 100$ GeV, one has $y_\B=\eta_{\rm B}/\eta_1\simeq 10^{-12}$, while $y_{\rm rec }\simeq 1$. Therefore, the above expression does not depend strongly on the magnetic field generation time. For the neutrino `magnetic' part, we find instead
\be
\Psist_{\rm mag}-\Phist_{\rm mag}=-\frac{12\, [(3 y^2 + 2y - (6 + 5 y) y_\nu ) (y-y_\nu)]}{y^4 (2 + y)^3} \, \frac{\Pi_\B}{x_1^2}~.
\label{PsimoinsPhinu}
\ee
Here we can see at play the compensating effect of neutrinos already proposed in~\cite{Finelli:2008xh,PFP,Shaw:2009nf,Kojima:2009gw} (see Fig.~\ref{fig:Pinu}). The contribution in Eq.~(\ref{PsimoinsPhiB}) that is not suppressed by $y_\B$ is exactly cancelled by the effect of the neutrino anisotropic stress. For the total Sachs Wolfe effect at leading order in the $x_1\ll 1$ expansion $\mathcal{O}(x_1^{-2})$, we find then
\be
\label{dT_neutrinos}
\left. \frac{\De T}{T}\right|_\B  \simeq -\frac{12 \, \Big[4 y_{\rm rec}^2 (y_B - 2 y_\nu) +y_{\rm rec} (4 y_B +5 y_\nu ^2 -8y_\nu)+ 6 y_\nu^2\Big]}{y_{\rm rec}^4 (2 + y_{\rm rec})^3} \frac{\Pi_\B}{x_1^2 }~.
\ee
This contribution is proportional to the time of generation of the magnetic field $y_\B\simeq 10^{-12}$ and to the time of neutrino decoupling, for which we have $y_\nu\simeq 10^{-6}$: it is therefore strongly suppressed. It actually only corresponds to the imprint of the magnetic field anisotropic stress from its time of generation $y_\B$ to the decoupling time of the neutrinos $y_\nu$. The subsequent magnetic contribution to the Sachs Wolfe effect, arising from $y_\nu$ up to recombination time, is cancelled by the free-streaming neutrinos. 

The dependence on $y$ and $y_\nu$ of \eqref{dT_neutrinos} follows from the particular form of the potentials (\ref{Phistress}) and (\ref{Psistress}), which in turns depends on the fit function $\pi_\nu(y)$ given in Eq.~(\ref{pinuapp}). Using this fit and setting $y_{\rm rec}=1$ gives the value
\be
\left. \frac{\De T}{T}\right|_\B  \simeq  7\cdot 10^{-6}~\frac{\Pi_\B}{x_1^2}~.
\label{dT_neutrinos_value}
\ee
This becomes of the order $\Pi_\B$ only at scales $k\lesssim 10^{-5}$ Mpc$^{-1}$: larger than the present horizon! A different fit could provide a different numerical value but would not alter the overall suppression effect due to the free-streaming neutrinos. For example, if we had assumed that $\pi_\nu(y)$ adjusts itself linearly to the compensating plateau $-3\Pi_\B/R_\nu$, instead of quadratically as chosen in \eqref{pinuapp}, the numerical value of the temperature anisotropy (\ref{dT_neutrinos_value}) would have changed by about 14\%. Even though the fit of \eqref{pinuapp} is better than a linear adjustment, conservatively the precision of our analytical analysis this should be taken to be of about 10\%. 

Note that, if we neglected neutrinos, the magnetic Sachs Wolfe effect would be given by \eqref{PsimoinsPhiB} taken at recombination time, $y=y_{\rm rec}$. This gives approximately 
\be
\left. \frac{\De T}{T}\right|_\B  \simeq \frac{20}{9}\,\frac{\Pi_\B}{x_1^2}~.
\label{dT_onlyB_value}
\ee
We will see in section \ref{nexttoleading} that this is of the same order of magnitude of the next-to-leading order contribution to the magnetic Sachs Wolfe. It would then have an observable effect on the CMB for a magnetic field of the order of the nanoGauss. It is only because of the equilibration among the magnetic and neutrino anisotropic stresses that this contribution is washed out, and not because it comes from a decaying mode (\cf discussion in the conclusion). If we had neglected neutrinos, as is usually done for the inflationary mode at large scales, we would still have found the correct amplitude for the magnetic Sachs Wolfe, but not the correct spectrum, as we now show. 

\subsection{CMB spectrum from the Sachs Wolfe effect at leading order}

With the above result we can give an estimate of the Sachs Wolfe contribution to the CMB spectrum. For the scope of this paper, and in order to give an interpretation of our result, we are merely interested in its general $\ell$-dependence. We do not aim at determining the amplitude of the magnetic contribution given in \eqref{dT_neutrinos_value} to the CMB spectrum at large scale in any details, since this is anyway completely unobservable. 

The CMB spectrum is given by (we use the notations of \cite{Hu:1997hp})
\be
C_\ell=\frac{2}{\pi} \frac{1}{(2\ell+1)^2} \int dk\,k^2\, \Theta_\ell (\eta_0,k)  \Theta_\ell^* (\eta_0,k)~,
\ee 
where at sufficiently large scales (\cf \eqref{SW} and the appendix of~\cite{giov}) one has
\be
\frac{\Theta_\ell (\eta_0,k)}{2\ell+1}\simeq g(\eta_{\rm rec})\left[\frac{D_{g\,\gamma}(k, \eta_{\rm rec})}{4} +\Psi(k, \eta_{\rm rec})-\Phi(k, \eta_{\rm rec})\right] j_\ell (k\eta_0)~,
\ee
where the integrated Sachs Wolfe effect is neglected, and $g(\eta_{\rm rec})$ denotes the visibility function. Since our treatment resides on the hypothesis of a causally created magnetic field, we assume that the magnetic and inflationary perturbations are not correlated. The magnetic contribution to the temperature anisotropy is then given by \eqref{dT_neutrinos}: since we are interested here only in the $\ell$-dependence of the CMB spectrum, for brevity we resume the amplitude of \eqref{dT_neutrinos} in a generic function $f(y_{\rm rec}, y_\nu,y_\B)$. We have then
\be
\frac{\Theta^\B_\ell(\eta_0,k)}{2\ell+1}\simeq g(\eta_{\rm rec}) \left[ f(y_{\rm rec}, y_\nu,y_\B) \, \frac{\Pi_\B}{x_1^2} \right] j_\ell(k\eta_0)~.
\label{ThetaB1}
\ee
The magnetic CMB spectrum becomes then
\be
C_\ell^\B\simeq \frac{2}{\pi}\, \frac{f^2(y_{\rm rec}, y_\nu,y_\B)}{\eta_1^4}\, g^2(\eta_{\rm rec}) \int dk\, \frac{|\Pi_\B(k)|^2}{k^2}\,j_\ell^2(k\eta_0)~,
\label{CellB1}
\ee
where $|\Pi_\B(k)|^2$ denotes the spectral amplitude of the magnetic field anisotropic stress
\be
\vev{\Pi_\B(\bk)\Pi_\B^*(\bq)}=(2\pi)^3\de(\bk-\bq)|\Pi_\B(k)|^2~.
\ee
This has been calculated in~\cite{Kahniashvili:2006hy,Finelli:2008xh,Shaw:2009nf,Brown:2005kr}, and it shares the same $k$-dependence as the spectral amplitude of the magnetic field energy density $\Om_\B$. For a causal magnetic field with $n\geq 2$ it is simply constant in $k$ up to the damping scale $k_D$, which we assume time-independent. We denote its amplitude by $\bar\Pi$, so that 
\be
|\Pi_\B(k)|^2=\bar{\Pi} \, \frac{\vev{B^2}^2}{\bar\rho_{\rm rad}^2} \, \frac{1}{k_D^3}~.
\label{PiBspec}
\ee
Substituting in \eqref{CellB1} and setting $x=k\eta_0$ we find then 
\bea
\hspace*{-0.8cm}\ell(\ell+1) C_\ell^\B &\simeq& \frac{2}{\pi} \, f^2(y_{\rm rec}, y_\nu,y_\B)\, g^2(\eta_{\rm rec}) \,\bar\Pi \, \frac{\vev{B^2}^2}{\bar\rho_{\rm rad}^2} \, \frac{\eta_0}{\eta_1} \, \frac{\ell(\ell+1)}{(\eta_1 k_D)^3} \int_0^{\eta_0k_D} dx \, \frac{j_\ell^2(x)}{x^2} \\
&\simeq & f^2(y_{\rm rec}, y_\nu,y_\B) \, g^2(\eta_{\rm rec}) \, \bar\Pi \, \frac{\vev{B^2}^2}{\bar\rho_{\rm rad}^2} \, \frac{\eta_0}{\eta_1} \,
\frac{1}{(\eta_1 k_D)^3} \, \frac{2\, \ell(\ell+1)}{8\ell^3+12\ell^2-2\ell-3}~,\nonumber
\eea
where for the last equality we have taken the limit $\eta_0k_D\gg 1$ (\cf \cite{Caprini:2009vk}). Therefore, we find a scaling for the CMB spectrum as $1/\ell$ for $\ell > 2$, as found in \cite{Kahniashvili:2006hy,Yamazaki:2008gr}. Note that, although this scaling might seem unusual, this contribution is of the same order of the standard inflationary one in the $x_1\ll 1$ expansion, \ie $a_1/x_1^2$ (\cf \eqref{dToverTinf}). Contrary to the magnetic field case for which the spectrum is flat (\cf \eqref{PiBspec}), the inflationary generated perturbations have an almost Harrison-Zeldovich spectrum corresponding to $|a_1(k)|^2 \propto k$, which leads to the usual flat CMB spectrum at large scales once inserted into \eqref{CellB1} (see Eqs.~(3.112) and (3.113) of \cite{libroruth}). 

\vspace*{1cm}

The $\ell$-dependence of the Sachs Wolfe effect coming from Eq.~(\ref{dT_neutrinos}) does not correspond to the result presented in \cite{Finelli:2008xh,PFP,Shaw:2009nf,yamafin}: in these works only the next-to-leading order contribution, constant in $x_1$, has been considered, leading to a CMB spectrum which behaves as $\ell(\ell+1)C_\ell^\B\propto \ell^2$. This is due to the different way of treating the initial conditions. Indeed, \cite{Finelli:2008xh,PFP,Shaw:2009nf,yamafin} only need to derive the initial conditions for the metric and fluid variables after neutrino decoupling, which are then inserted into the Boltzmann code. The contribution to the temperature anisotropy of the magnetic field anisotropic stress from its time of generation $y_\B$ to the decoupling time of the neutrinos $y_\nu$ is therefore absent. Moreover, the time evolution of the metric and fluid variables at order $\mathcal{O}(1/x_1^2)$ which we have obtained above has been identified with a decaying mode, and therefore neglected since the beginning (\cf discussion in section \ref{sec:conc}). When one derives the initial conditions in the syncronous gauge, as done in \cite{Finelli:2008xh,PFP,Shaw:2009nf,yamafin}, it results that the only way to avoid the long wavelength mode at $\mathcal{O}(1/x_1^2)$  and select the constant mode at order $\mathcal{O}(x_1^0)$ is to solve the Einstein equations setting the neutrino anisotropic stress to be constant in time, and exactly equal and opposite to the magnetic field one. This is why in \cite{Finelli:2008xh,PFP,Shaw:2009nf,yamafin} the anisotropic stress of the magnetic field is compensated by the one of the neutrinos since the beginning, directly in the initial conditions. In our derivation, the anisotropic stress of the neutrinos is instead time dependent: we derive its behaviour in time in appendix \ref{app:pinu} and then insert it in Einstein's equations by means of the analytical fit given in \eqref{pinuapp}. Therefore, here we obtain the compensation effect dynamically, without having to insert it directly in the initial conditions.

Within our treatment, we can somehow mimic the result of \cite{Finelli:2008xh,PFP,Shaw:2009nf,yamafin} if we set  $\pi_\nu=-3\Pi_\B /R_\nu$ and $y_\nu=y_\B$\footnote{Note that the condition $y_\nu=y_\B$ is necessary if the neutrino anisotropic stress adjusts itself instantaneously to the magnetic field one. A sudden jump in the neutrinos anisotropic stress induces in fact a discontinuity in the total energy momentum tensor, and we can only arrange this by means of including it in the matching at $y_\B$.}. In this case, the source term $\pi_{\rm F}=R_\nu\pi_\nu$ in Eq.~(\ref{Dddot}) exactly cancels the magnetic source term $\Pi_\B$. We can then repeat what done in Sec.~\ref{without_n}, where the source $S_\B$ of \eqref{SB} is now only given by the $\Om_\B$-part. The matching conditions are different and the solution for $D$ becomes simply
\be
D^+(y)=-\frac{\Om_B}{(1+y)^2}~.
\ee
The source term in \eqref{dToverTBcomplete} then vanishes at leading order $\mathcal{O}(1/x_1^2)$ and therefore the Sachs Wolfe contribution is exactly zero at this order, as found in \cite{giov}. 

Our analytical solution~(\ref{dT_neutrinos}) shows that the period between the generation of the magnetic field and the time of neutrino decoupling would leave an imprint on the Sachs Wolfe that is diverging at large scales. However, since this imprint is suppressed by $y_\B$ and $y_\nu$ as a consequence of the neutrino compensating effect, it leaves no observable impact. Therefore, we now proceed to compute analytically the next order contribution $\mO(x_1^0)$ in the Sachs Wolfe. 

\section{Next-to-leading order solutions at large scales $x_1\ll 1$}
\label{nexttoleading}

The temperature anisotropy at constant order in $x_1$ is given by Eq.~(\ref{dToverTBcomplete}), where we now need to take into account the next-to-leading order $\mathcal{O}(x_1^0)$ in $\Psi^{\B}-\Phi^{\B}$ and $\Psist_{\rm mag}-\Phist_{\rm mag}$. The Lorentz force term, which could be neglected in the above result, must be taken into account now, since it is constant in $x_1$. The integrals are one order higher with respect to their integrands, so to get the contribution constant in $x_1$ it is enough to keep the order $\mathcal{O}(x_1^{-2})$ in the sources $\Psi^{\B}-\Phi^{\B}$ and $\Psist_{\rm mag}-\Phist_{\rm mag}$.  

The easiest way to compute the order $\mathcal{O}(x_1^0)$ in $\Phi$ and $\Psi$ is to use the curvature perturbation $\zeta$. Indeed, starting from the solution given in Eq. (\ref{Dstress}) for $D^{\rm fin}$, we can calculate the order $\mO(x_1^0)$ in $\zeta$. Once this is known, we can use definition (\ref{zeta}) to integrate for $\Phi$ at the same order. 

Deriving \eqref{zeta}, using definition (\ref{Einstein3}) and the momentum conservation equation (\ref{consmom}), we find the evolution equation for $\zeta$ 
\be
\frac{d\zeta}{dy}=\frac{\HH \eta_1}{1+w}\left[ c_s^2 D +\frac{\Om_\B-2\Pi_\B}{3(1+a)} -\frac{2}{3}w\pi_{\rm F} \right]~.
\label{zetadot}
\ee
Since the leading order in $D$ is $\mathcal{O}(x_1^0)$, the source term in the above equation is also of the same order: therefore, we can solve the equation to find $\zeta$ at order constant in $x_1$. 

In solving \eqref{zetadot} we concentrate here only on the magnetic part and neglect the inflationary one, which is not relevant for the scope of the paper. This means that, for $y_\B<y<y_\nu$, the source is given by the variable $D^\B$ (\cf \eqref{Dplusy}). Furthermore, we have $\pi_{\rm F}=0$ 
and we solve Eq.~(\ref{zetadot}) imposing the continuity of $\zeta$ at $y_\B$: for the magnetic part this means $\zeta^\B(y_\B)=0$, which is equivalent to the matching conditions. In the limit $y_\B<y\ll 1$, the solution reads
\be
\left. \zeta^\B (y_\B<y<y_\nu)\right|_{\mO(x_1^0)} \simeq \frac{\Pi_\B}{2}\left[1-\frac{y_\B}{y} -2\log \left(\frac{y}{y_\B}\right) \right]~.
\label{zetacorrect}
\ee 
This result is consistent with what given in Eq.~(84) of \cite{Shaw:2009nf}: the curvature is sourced only by the magnetic field anisotropic stress. The reason for this is that in Eq.~(\ref{zetadot}), initially, the fluid energy density $D^\B$ is compensated by the magnetic field energy density $\Om_\B$  because of the matching condition (\ref{matchD}) and because $c_s^2\simeq 1/3$ and $\pi_{\rm F}=0$. Therefore, the only possible source for $\zeta$ is proportional to $\Pi_\B$. With the evolution, the compensation of $D^\B$ is no longer perfect, and $D^\B$ becomes proportional to $\Pi_\B$, Eq.~(\ref{Dplusy}). Therefore, the curvature only depends on $\Pi_\B$. 

After neutrino decoupling, we use our approximation (\ref{pinuapp}) for $\pi_\nu$ and insert it in Eq.~(\ref{zetadot}). The rest of the source is now logically given by $D^\B+\Dst_{\rm mag}$. As before, we impose the continuity of $\zeta$ at $y_\nu$ using the solution determined previously, $\zeta(y_\nu)=\zeta^\B(y_\nu)$. We find, for $y_\B<y_\nu<y\ll 1$,
\be
\left.\zeta(y_\nu<y<1)\right|_{\mO(x_1^0)} = \zeta^\B+\zeta^{\rm as}_{\rm mag}  \simeq \frac{\Pi_B}{2} \left[ -1 + 2 \, \frac{y_\nu}{y} -2 \log \left(\frac{y_\nu}{y_B}\right) \right] ~, 
\label{zetaynusmall}
\ee
which reduces to \eqref{zetacorrect} for $y=y_\nu$. The curvature stops growing after neutrino decoupling, as already pointed out in \cite{Shaw:2009nf}, since the source of Eq.~(\ref{zetadot}) is fully compensated: the neutrino anisotropic stress compensates the magnetic field one, and $\Dst_{\rm mag}$ compensates the part proportional to $\Pi_\B$ in $D^\B$ which acted as a source for $y<y_\nu$ (\cf Eqs.~(\ref{Dplusy}) and (\ref{Dstress})). In the limit $y \gg 1>y_\nu$ we find instead
\be
\left.\zeta(y>1>y_\nu)\right|_{\mO(x_1^0)} = \zeta^\B+\zeta^{\rm as}_{\rm mag} \simeq  \frac{\Om_B}{4} + \Pi_B \left[ -\frac{1}{2} -  \log \left(\frac{y_\nu}{y_B}\right) \right] ~. 
\label{zetaynu}
\ee
This shows that after equality, while the neutrinos are still a relativistic component and therefore continue to compensate the magnetic field anisotropic stress, the fluid is no longer relativistic, and no compensation is possible among its energy density perturbation and the magnetic energy density: therefore, the source of (\ref{zetadot}) is now given by $\Omega_\B$ which is no longer compensated, and the curvature grows. Comparing the above equation with Eq.~(86) of \cite{Shaw:2009nf}, we see that the mode sourced by the magnetic anisotropic stress is very well approximated by our analytical fit: the only difference among the two equations is the term $-1/2$ in \eqref{zetaynu} which becomes $1-5/(8R_\nu)\simeq 0.56$ in Eq.~(86) of \cite{Shaw:2009nf}. In our approach, however, we do not distinguish among the `passive' and `compensated' modes (see \cite{Shaw:2009nf}): therefore, in \eqref{zetaynu} we get an additional contribution in the curvature proportional to $\Om_\B$.  

Note that the curvature evolves with time at next-to-leading order in $x_1\ll 1$, but this does not affect curvature conservation at large scales. First of all, the magnetic field contribution at order $\mathcal{O}(1/x_1^{2})$ remains zero, as given in Eq.~(\ref{zetanu}). Moreover, at next-to-leading order, the effect of the magnetic field on the curvature spectrum is constant in $x_1$: $|\zeta_\B(k)|^2\propto x_1^0$, since the magnetic field anisotropic stress spectrum and the magnetic field energy density spectrum are flat for a causally generated field (up to the damping scale $k_D$, see Eqs.~(\ref{PiBspec}) and (\ref{OmBk})). On the contrary, the inflationary contribution to the curvature spectrum scales as $|\zeta_{\rm inf}(k)|^2\propto x_1^{-3}$, and is therefore the dominant contribution at large scales. Hence, as already seen in section~\ref{with_n}, the curvature is still conserved at large scales, even in the presence of a magnetic field.

We can now use the solution for $\zeta$ at order $\mO(x_1^0)$ to compute $\Phi$ and $\Psi$ at the same order. Using Einstein's equation~(\ref{Einstein2}), Eq.~(\ref{zeta}) can be rewritten as an evolution equation for $\Phi$
\be
\label{phizeta}
\frac{d\Phi}{dy}+\frac{\HH\eta_1(5+3w)}{2}\Phi=-3w\HH\eta_1\left(\frac{\HH}{k} \right)^2\Big[R_\nu\pi_\nu+3\Pi_\B \Big] -\frac{3\HH\eta_1(1+w)}{2}\zeta~.
\ee
Before neutrino decoupling, $\pi_\nu=0$ and the source-term at order $\mO(x_1^0)$ is given only by the term proportional to $\zeta$ ($\Pi_\B$ contributes at order $\mO(x_1^{-2})$). In order to determine the magnetic contribution to the Bardeen potential $\Phi^\B$, we therefore solve the above equation with $\zeta^\B(y_\B<y<y_\nu)$  as source (given in \eqref{zetacorrect} in the limit $y\ll 1$). The initial condition for (\ref{phizeta}) in this case is found by imposing the continuity of $\Phi^\B$ at $y_\B$, $\Phi^\B(y_\B)=0$. Again in the limit $y_\B<y<y_\nu \ll 1$, the solution reads
\be
\left. \Phi^\B(y)\right|_{\mO(x_1^0)} = \left[ -\frac{5}{9} + \frac{2}{3} \log\left(\frac{y}{y_\B}\right) +\frac{y_B}{2 y}\right] \Pi_B
- \left[ \frac{3}{16} \Om_\B + \frac{95}{144} \Pi_\B + \log\left(\frac{y}{y_\B}\right) \Pi_\B \right] y~.
\label{PhiBnext}
\ee
In this expansion we have kept the term proportional to $\Om_\B \, y$, since it corresponds to the solution given in the analytical estimate of~\cite{giov}, Eq.~(3.20), where however the magnetic anisotropic stress has been neglected\footnote{In the same way, ignoring the free-streaming neutrinos and taking the limit $y\gg 1$, we would get
\bea
\left. \zeta^\B(y)\right|_{\mO(x_1^0)} = \frac{\Om_\B}{4}+\Pi_\B\left[ 2 - \log\left( \frac{2}{y_\B}\right)\right], ~~~~
\left. \Phi^\B(y)\right|_{\mO(x_1^0)} = -\frac{3}{20} \Om_\B - \left[ \frac{6}{5} -\frac{3}{5}\log\left(\frac{2}{y_\B}\right)  \right] \, \Pi_\B\,, \nonumber
\eea
where again the term proportional to $\Om_\B$ in $\Phi^\B$ corresponds to the solution given in Eq.~(3.20) of \cite{giov}.}.

Solving for $\Phi$ after neutrino decoupling is more involved. Indeed, looking at (\ref{phizeta}) we see that the neutrino anisotropic stress at order $\mO(x_1^2)$ contributes to the source term at order $\mO(x_1^0)$. However, we cannot determine analytically $\pi_\nu$ at order $\mO(x_1^2)$: this would require to redo the calculation presented in appendix \ref{app:pinu} at the following order, and therefore to solve the whole system of coupled differential equations for the metric and the individual fluid components. This cannot be done analytically and is beyond the scope of this paper. We can however find an approximate solution by taking for $\pi_\nu$ the following Ansatz 
\be
\left. \pi_\nu(y)\right|_{\mO(x_1^2)}=\Big(d_1\Om_\B+d_2\Pi_\B\Big)(y-y_\nu)^2 x_1^2~,
\label{pinuOx12}
\ee
with $d_1$ and $d_2$ two arbitrary constants that we need to determine. This Ansatz has been chosen because it is consistent with the initial conditions of \cite{Finelli:2008xh,PFP,Shaw:2009nf}. It allows us to determine the first part of the source in (\ref{phizeta}) as a function of the constants $d_1$ and $d_2$. The remaining part of the source is given by the curvature perturbation at $y>y_\nu$, which we have determined previously and written in Eqs.~ (\ref{zetaynusmall}) and (\ref{zetaynu}) respectively in the limits $y\ll 1$ and $y\gg 1$. 

Knowing the source, we can solve \eqref{phizeta} at order $\mO(x_1^0)$ and determine $\Phi$ at this order as a function of $d_1$ and $d_2$. Note that $\Phi$ in this case corresponds to $\Phi^\B+\Phist_{\rm mag}$, \ie the magnetic part plus the neutrino magnetic one. The constant of integration is fixed by imposing that $\Phi$ is continuous at $y_\nu$, i.e. $\Phist_{\rm mag}(y_\nu)=\Phi^\B(y_\nu)$ as derived above. Once $\Phi$ is known, using Einstein's equations we can then derive $\Psi\,,~D$ and $V$ after neutrino decoupling as functions of $d_1$ and $d_2$. The constants $d_1$ and $d_2$ can then be determined solving the system of equations for the individual fluid components: in particular, the energy and momentum conservation equations for neutrinos. The derivation is presented in appendix~\ref{app:pinunext}. We find then, in the limit $y_\nu\ll 1$ 
\be
\left. \pi_\nu(y)\right|_{\mO(x_1)^2}=\frac{
\left[ 6-4R_\nu + 8 R_\nu \log \left( y_\B/y_\nu \right) \right] \,\Pi_\B 
-3 \, \Om_\B\, R_\nu }{2\,R_\nu(15+4 R_\nu)}  \,(y-y_\nu)^2\,x_1^2~,
\label{pinunext}
\ee
while for $\Phi=\Phi^\B+\Phist_{\rm mag}$ we find, for $y_\nu <y \ll 1$
\be
\left. \Phi(y)\right |_{\mO(x_1^2)}= \frac{R_\nu \, \Om_\B+ \left[ 4 (2+ R_\nu) + 4 (5 +2 R_\nu) \log(y_\nu/y_\B)  \right]
 \Pi_\B }{2\,(15+4 R_\nu)}~,
 \label{phix12ysmall}
\ee
and for $y\gg 1$
\be
\left. \Phi(y)\right|_{\mO(x_1^2)}=-\frac{3}{20} \Om_\B + \left[ \frac{3}{10} + \frac{3}{5} \log\left( \frac{y_\nu}{y_\B} \right) \right]\,\Pi_\B~.
\label{phix12ybig}
\ee
The full time dependence of the magnetic Bardeen potential $\Phi(y)$ is rather complicated, so we do not write it here. It is shown in Fig.~\ref{figphi} where the transition from the radiation to the matter era is apparent. Note that, to solve both for the curvature and the Bardeen potential we made use of our fit for the neutrino anisotropic stress at leading order in $k\eta_1\ll 1$, see Fig.~\ref{fig:Pinu}. This allows us to find the complete solutions for the metric perturbations across the radiation-matter transition; however, it also causes minor differences among the early time limit of our analytical solutions and the corresponding initial conditions given in \cite{Shaw:2009nf} (for example, \eqref{phix12ysmall} slightly differs in the $R_\nu$ dependence from the one that we could derive combining the passive and active modes given in the appendix of \cite{Shaw:2009nf}).

\FIGURE{\epsfig{file=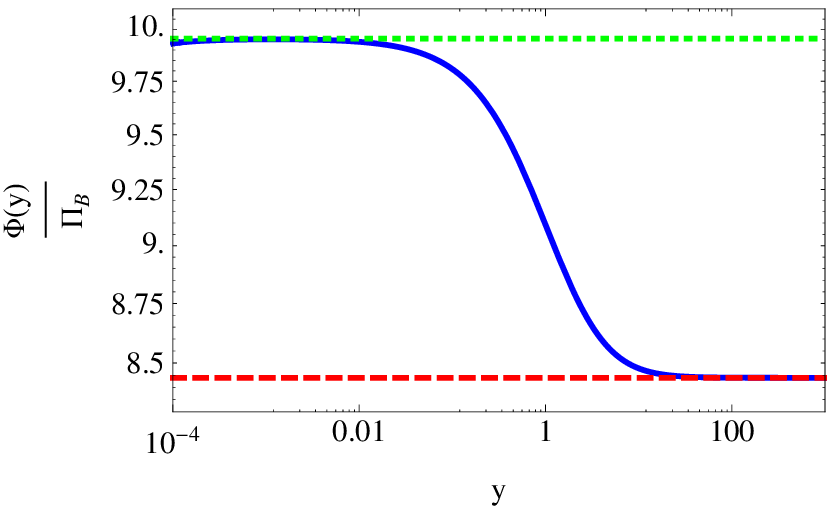,width=10cm} 
        \caption{Solid, blue: the magnetic contribution to the Bardeen potential at next-to-leading order $\left.\Phi(y)\right|_{\mO(x_1^2)}$, across the radiation-matter transition, together with its approximations at early and late time: \eqref{phix12ysmall}, dotted, green and \eqref{phix12ybig}, dashed, red. In the plot, the magnetic energy density and magnetic anisotropic stress have comparable amplitude, and we have normalised the amplitude of $\Phi$ to $\Pi_\B$.}
	\label{figphi}}

From the above solution for $\Phi$ we can compute $\Psi$, and in turns the Sachs Wolfe at order $\mO(x_1^0)$. This involves the computation of all the terms of \eqref{dToverTBcomplete}, and its expression as a function of $y$ is again rather complicated. We write it here for $y=y_{\rm rec}\simeq 1$
\bea
\left. \frac{\De T}{T}\right|_{\B\,\mO(x_1^0)}&\simeq&\frac{1}{540 (15 + 4R_\nu)(R_\nu-1)}\Bigg\{ \Big[495 +2R_\nu(811+224R_\nu)\Big] \, \Omega_\B  \\
& + &\left[ -2960-2R_\nu(977+108R_\nu) +  4(R_\nu-1) (505 + 108 R_\nu) \log\left(\frac{y_\B}{y_\nu}\right) \right] \Pi_\B \Bigg\} \nonumber \\
& \simeq & -0.2 \, \Om_\B - 2.7 \, \Pi_\B~, \nonumber 
\eea 
where in the last line we have set $R_\nu= 0.4$, $y_\nu=10^{-6}$ and $y_\B=10^{-12}$. Since the temperature anisotropy in the above equation is independent of $k$, it gives rise to a CMB spectrum scaling as $\ell(\ell+1)\,C^\B_\ell \propto \ell^2$, as we now demonstrate. 

The equivalent of Eq.~(\ref{ThetaB1}) in this case is
 \be
\frac{\Theta^\B_\ell(\eta_0,k)}{2\ell+1}\simeq g(\eta_{\rm rec}) \left[ -0.2\, \Om_\B-2.7\, \Pi_\B \right] j_\ell(k\eta_0)~,
\ee
and the magnetic CMB spectrum becomes then (\cf \eqref{CellB1})
\be
C_\ell^\B\simeq \frac{2}{\pi} \, g^2(\eta_{\rm rec}) \int dk\, k^2\, \left[  0.04\, |\Om_\B(k)|^2+ 7.29 \, |\Pi_\B(k)|^2+ 0.54 \, |C_\B(k)|^2  \right]\,j_\ell^2(k\eta_0)~,
\label{CellB2}
\ee
where $|\Pi_\B(k)|^2$ is given in \eqref{PiBspec}, $|\Om_\B(k)|^2$ denotes the spectral amplitude of the magnetic field energy density,
\be
\vev{\Om_\B(\bk)\Om_\B^*(\bq)}=(2\pi)^3\de(\bk-\bq)|\Om_\B(k)|^2~,
\ee 
and $|C_\B(k)|^2$ denotes the spectral amplitude of the cross-correlation
\be
\vev{\Om_\B(\bk)\Pi_\B^*(\bq)}=(2\pi)^3\de(\bk-\bq)|C_\B(k)|^2~.
\ee 
The energy density spectrum and the cross-correlation one have been calculated in~\cite{Kahniashvili:2006hy,Finelli:2008xh,Shaw:2009nf}, and they all share the same $k$-dependence: similarly to \eqref{PiBspec}, we set then
\be
|\Om_\B(k)|^2=\bar{\Om} \, \frac{\vev{B^2}^2}{\bar\rho_{\rm rad}^2} \, \frac{1}{k_D^3} \label{OmBk} ~~~~~~~~~~
|C_\B(k)|^2=\bar{C} \, \frac{\vev{B^2}^2}{\bar\rho_{\rm rad}^2} \, \frac{1}{k_D^3}~.
\ee 
Substituting the above equations in \eqref{CellB2}, we find
\bea
\hspace*{-1cm}\ell(\ell+1) C_\ell^\B &\simeq& \frac{2}{\pi} \, g^2(\eta_{\rm rec}) \,\left[ 0.04 \, \bar\Om +7.29  \, \bar\Pi +0.54  \, \bar{C} \right]\, \frac{\vev{B^2}^2}{\bar\rho_{\rm rad}^2} \, \frac{\ell(\ell+1)}{(\eta_0 k_D)^3} \,  \int_0^{\eta_0k_D} dx \, x^2\, j_\ell^2(x)  \nonumber \\ 
&\simeq & g^2(\eta_{\rm rec}) \, \left[ 0.04 \, \bar\Om +7.29 \,  \bar\Pi +0.54  \, \bar{C} \right] \, \frac{\vev{B^2}^2}{\bar\rho_{\rm rad}^2} \, \frac{\ell(\ell+1)}{\pi\,(\eta_0 k_D)^2}~, 
\eea
where for the last equality we have used approximation (A2) of \cite{Caprini:2009vk}. Therefore, we confirm that the next-to-leading order contribution to the Sachs Wolfe effect behaves as found in~\cite{Finelli:2008xh,PFP,Shaw:2009nf,yamafin}.

\section{Conclusions}
\label{sec:conc}

In this work we present an analytical computation of the Sachs Wolfe effect induced by a primordial magnetic field. We have restricted our analysis to a magnetic field generated by a causal process, which acts on a time scale shorter than the Hubble time at generation, such as a first order phase transition in the early universe. The reason is that, under this hypothesis, the initial conditions for the metric and fluid variables are determined unambiguously by imposing conservation of the total energy momentum tensor across the transition, just as in the topological defects case \cite{Deruelle:1997py,Uzan:1998rt}. After setting the initial conditions with this matching, we have solved Einstein's and conservation equations analytically and computed the magnetic effect on the gauge invariant metric and fluid perturbations in the large scales limit $k\eta_1\ll 1$, \ie for scales larger than the horizon at recombination. Using these solutions, we have then determined the Sachs Wolfe effect. This is sourced only by the magnetic anisotropic stress: the temperature perturbation behaves as $\Delta T/T\propto \Pi_\B/(k\eta_1)^2$ at leading order in the $k\eta_1\ll 1$ expansion. A scaling of the CMB spectrum as $\ell(\ell+1)C^\B_\ell\propto \ell^{-1}$ trivially follows.

However, as already pointed out in \cite{Shaw:2009nf,Kojima:2009gw}, the magnetic anisotropic stress is quickly compensated by the neutrino one, once neutrinos decouple from the primordial fluid. We have verified this compensation analytically, by solving the neutrino evolution equations in the limit $k\eta_1\ll 1$, where we can consistently neglect multipoles higher than the anisotropic stress. Consequently, the contribution to the Sachs Wolfe at leading order in $k\eta_1\ll 1$ gets strongly suppressed after neutrino decoupling, and the next-to-leading order in $k\eta_1\ll 1$ becomes the relevant one: the one  which could in principle provide an observable effect in the CMB. In order to derive the magnetic Sachs Wolfe effect at next-to-leading order in $k\eta_1\ll 1$ we had to compute the metric and fluid variables consistently at this order. This can be done analytically by solving the evolution equation for the curvature perturbation, and leads to a temperature anisotropy which is sourced both by the magnetic anisotropic stress and the magnetic energy density: we found the approximate result $\Delta T/T\simeq -0.2 \, \Om_\B -2.7 \,\Pi_\B$. The new scaling of the CMB spectrum,  arising from this contribution, takes the form $\ell(\ell+1)C^\B_\ell\propto \ell^{2}$. 

The matching conditions which we used to determine the initial conditions for the magnetic field perturbations imply that the curvature perturbation is continuous across magnetic field generation time. The subsequent evolution of the curvature perturbation is then completely determined. We found that, at leading order in $k\eta_1\ll 1$, the magnetic field does not contribute to the curvature, which is therefore conserved at large scales and given only by the inflationary mode (for which we have imposed adiabatic initial conditions). At next-to-leading order in $k\eta_1\ll 1$, the curvature perturbation is instead affected by the presence of the magnetic field. As already found in \cite{Shaw:2009nf}, we confirm that, between the magnetic field generation time and neutrino decoupling, the curvature is sourced by the magnetic anisotropic stress and therefore grows (starting from zero at magnetic field generation time). After neutrino decoupling, when the magnetic anisotropic stress gets compensated, the part sourced by it remains constant, and the curvature perturbation gets a contribution from the magnetic energy density which is no longer compensated. 

Our analysis provides a clarification of the results present in the literature on the Sachs Wolfe effect arising from a primordial magnetic field. The magnetically induced CMB anisotropies have been studied in details in the literature by numerical integration of Boltzmann codes \cite{Finelli:2008xh,PFP,Shaw:2009nf,yamafin}, and in most of the cases the large scale CMB spectrum behaves as  $\ell(\ell+1)C^\B_\ell\propto \ell^{2}$, corresponding to a temperature anisotropy which does not depend on wavenumber. This is a consequence of the choice of the initial conditions. A Boltzmann code needs the initial conditions to be specified after neutrino decoupling, so the contribution to the Sachs Wolfe effect arising from the period of time between the magnetic field generation and the neutrino decoupling has been neglected in \cite{Finelli:2008xh,PFP,Shaw:2009nf,yamafin}, and the magnetic field anisotropic stress is exactly compensated by the free-streaming neutrino one since the beginning. This sets to zero the temperature anisotropy coming from the metric and fluid perturbations at leading order in $k\eta_1\ll 1$ to start with. This choice of the initial conditions has been motivated by an analogy with the inflationary case, for which one can identify a growing and a decaying mode, and consistently neglect the decaying one. In fact, the inflationary solution for the Bardeen potential is, for $y\ll 1$
\be
\Phi^- (y) \simeq  \frac{3}{4\,y^3}\,\frac{b_1}{x_1^2}+\frac{20}{3}\,\frac{a_1}{x_1^2}= \frac{3}{4}\, \frac{\eta_1}{\eta}\,\frac{b_1}{(k\eta)^2}+\frac{20}{3}\left(\frac{\eta}{\eta_1}\right)^2 \frac{a_1}{(k\eta)^2}~,
\ee
where the mode proportional to $b_1$ is decaying with time, while the one proportional to $a_1$ is constant. One expects that from inflation the two modes are generated with the same amplitude: consequently, $b_1$ must be much smaller than $a_1$ since, at very early times (\eg reheating), $\eta_1/\eta_{\rm reh}\gg 1$ \cite{libroruth}. The Sachs Wolfe effect from both modes at leading order in $k\eta_1\ll 1$ is given by (\cf also \eqref{dToverTinf})
\be
\left. \frac{\De T}{T}\right|_{\rm inf} \simeq -\frac{12(1+y_{\rm rec})}{y_{\rm rec}^3(2+y_{\rm rec})^3} \, \frac{b_1}{x_1^2}-\frac{2(40+40y_{\rm rec}+18y_{\rm rec}^2+3y_{\rm rec}^3)}{3(2+y_{\rm rec})^3}\, \frac{a_1}{x_1^2}~. 
\ee
The two modes have the same dependence on $k$, and about the same amplitude for $y_{\rm rec}\simeq 1$: therefore, knowing that $b_1\ll a_1$, one can neglect the contribution of the decaying mode (for an analysis of the effect of the inflationary decaying mode on the CMB, see \cite{Amendola:2004rt}).  

The situation for the metric perturbation induced by the magnetic field is a bit different. In this case, the solution at leading and next-to-leading order in $k\eta_1\ll 1$ reads, for $y\ll 1$ (\cf also \eqref{Phiplus} and \eqref{PhiBnext})
\bea
\Phi^B (y) & \simeq & -\frac{3}{y^2}\left(1-\frac{y_\B}{y}\right) \frac{\Pi_\B}{x_1^2}+\left[ -\frac{5}{9}+\frac{2}{3}\log\left(\frac{y}{y_\B}\right)+\frac{y_\B}{2y}\right]\Pi_\B \\
& = & -3\left(1-\frac{\eta_\B}{\eta}\right) \frac{\Pi_\B}{(k\eta)^2}+\left[ -\frac{5}{9}+\frac{2}{3}\log\left(\frac{\eta}{\eta_\B}\right)+\frac{\eta_\B}{2\eta}\right]\Pi_\B~. \nonumber
\eea
The term at leading order in $k\eta_1\ll 1$ is decaying with time as $y^{-2}$, whereas the next-to-leading order is growing logarithmically. One could therefore be tempted to neglect the first term on the right hand-side of the above equation. However, the key point here is that, contrary to the inflationary case, the two modes not only have no independent amplitudes (they are both proportional to $\Pi_\B$), but also do not share the same $k$-dependence. Consequently, we see that the `decaying' mode always has the same amplitude of the `growing' one at horizon crossing $k\eta\simeq 1$, proportional to a few $\Pi_\B$. Since the CMB in general selects the contribution of modes which cross the horizon at recombination time $\eta_1$ (the first peak), this means that the two modes can have the same impact on the CMB for $k\eta_1\simeq 1$. Moreover, for the Sachs Wolfe effect, which selects modes still outside the horizon at recombination, we see that the `decaying' mode could even dominate.  

However, the analyses of \cite{Finelli:2008xh,PFP,Shaw:2009nf,yamafin} do get the correct result: as we have seen, this is due to the compensating effect of the neutrinos, which exactly cancels $\Pi_\B$ at leading order in $k\eta_1\ll 1$, strongly suppressing the amplitude of the `decaying' mode. The relevant contribution to the CMB anisotropies is therefore caused by the `growing' mode\footnote{Note that, at next-to-leading order in $k\eta_1\ll 1$ the cancellation from the neutrinos is not active: \cf the different amplitudes of \eqref{pinuapp} and \eqref{pinunext}}. The correct CMB spectrum at large scales from the Sachs Wolfe effect from a causally generated magnetic field with spectral index $n\geq 2$ is therefore $\ell(\ell+1)C^\B_\ell\propto \ell^{2}$. Note however, that in order for this effect to be observable, the magnetic field should have an amplitude of at least a few nanoGauss: causal magnetic fields generated before Nucleosynthesis are unfortunately strongly constrained to amplitudes far too small to leave any observable effect in the CMB \cite{Caprini:2001nb}.

\acknowledgments
We are grateful to Francis Bernardeau, Ruth Durrer, Fabio Finelli, Kazuya Koyama, Antony Lewis, Roy Maartens, Daniela Paoletti, Cyril Pitrou, Richard Shaw, Kandaswamy Subramanian, and Filippo Vernizzi for very useful discussions. 

\appendix

\section{The metric and fluid perturbations at leading order}
\label{app1}

In this appendix we present the solutions for the metric perturbations $\Phi$ and $\Psi$, and the velocity $V$ before and after the magnetic field generation, which are calculated via the matching procedure explained in section \ref{init_cond}. 

Before neutrino decoupling, from the expression of the density variable $D^+$ given in \eqref{Dplusy} one gets the Bardeen potential $\Phi$ from Einstein's equation (\ref{Einstein1})
\be
\Phi^+(y)=\Phi^-(y)+\Phi^\B(y)=\frac{2(8+3y)(10+10y+3y^2)}{3(2+y)^3}\,\frac{a_1}{x_1^2}-\frac{24(1+y)(y-y_\B)}{y^3(2+y)^3}\,\frac{\Pi_\B}{x_1^2}~.
\label{Phiplus}
\ee
The Bardeen potential $\Psi$ is obtained from Einstein's equation (\ref{Einstein2})
\be
\Psi^+(y)=\Psi^-(y)+\Psi^\B(y)=-\frac{2(8+3y)(10+10y+3y^2)}{3(2+y)^3}\,\frac{a_1}{x_1^2}+\frac{12(y^2-2y y_\B-2y_\B)}{y^3(2+y)^3}\,\frac{\Pi_\B}{x_1^2}~.
\label{Psiplus}
\ee
Note that the inflationary part is just opposite to $\Phi^-$. The velocity perturbation $V$ is obtained from Einstein's equation (\ref{Einstein3})
\be
V^+(y)=V^-(y)+V^\B(y)=-\frac{2 y (20+15y+3y^2)}{3(2+y)^2}\,\frac{a_1}{x_1}-\frac{12(y-y_\B)}{y^2(2+y)^2}\,\frac{\Pi_\B}{x_1}~.
\label{Vplus}
\ee

After neutrino decoupling, to the above expressions one needs to add the neutrino contribution generated by the neutrino anisotropic stress. This can be calculated again from Einstein's equations starting from the additional contribution to the density variable given in \eqref{Dstress}. One gets then
\bea 
\Phist&=& \Phist_{\rm inf}+\Phist_\B \label{Phistress} \\
&=& \frac{ 160\, R_\nu}{3  (\rnu)}  \frac{(1 + y)  (2 y + y_\nu) (y - y_\nu)^2 }{  y^3 (2 + y)^3} \,\frac{a_1}{x_1^2}+
\frac{ 24  (1 + y) (y - y_\nu)^2}{y^4 (2 + y)^3} \,\frac{\Pi_\B}{x_1^2}~, \nonumber \\
\Psist&=& \Psist_{\rm inf}+\Psist_\B \label{Psistress} \\
&=& \frac{ 160\, R_\nu}{3  (\rnu)}  \frac{y^3 (4 + y) - 3 y^2 y_\nu - (1 + y) y_\nu^3}{ y^3 (2 + y)^3}\,\frac{a_1}{x_1^2}-
\frac{12 (y - y_\nu) (y^2 - 4 y_\nu - 3 y y_\nu)}{y^4 (2 + y)^3}\,\frac{\Pi_\B}{x_1^2}~, \nonumber \\
\Vst&=& \Vst_{\rm inf}+\Vst_\B \label{Vstress} \\
&=& \frac{ 160\, R_\nu}{3  (\rnu)}  \frac{ (2 y + y_\nu) (y - y_\nu)^2}{ y^2 (2 + y)^2}\,\frac{a_1}{x_1}+\frac{12 (y - y_\nu)^2}{y^3 (2 + y)^2}\,\frac{\Pi_\B}{x_1}~. \nonumber
\eea
The above expressions depend on the particular form of the time dependence of the neutrino anisotropic stress that we choose, given in \eqref{pinuapp}.

\section{Resolution for the neutrino anisotropic stress at leading order}
\label{app:pinu}

We present here the derivation of the neutrino anisotropic stress at leading order $\mO(x_1^0)$. We follow the method of \cite{Shaw:2009nf}. In order to proceed analytically, we solve in the radiation era. The anisotropic stress so derived will be used to calculate the Sachs Wolfe effect and consequently it is assumed to be valid up to recombination time. Following \cite{Shaw:2009nf}, we derive a fourth-order differential equation for $\pi_\nu$ by combining Einstein's equations and the Boltzmann hierarchy 
\bea
D_{s\, \nu}'(y)+\frac{4x_1}{3}V_{\nu}(y)&=&-4\Phi'(y)~,  \label{Dsnu}\\
\frac{4}{x_1}V_\nu'(y)+\frac{2}{3}\pi_\nu(y)-D_{s\, \nu}(y)&=&4\Psi(y)~,\label{Vnu}\\
\frac{5}{3 x_1}\pi_\nu'(y)+\frac{3}{7}F_3(y)-\frac{8}{3}V_\nu(y)&=&0~,\label{pinu}\\
\frac{1}{x_1}F_\ell'(y) +\frac{\ell+1}{2\ell+3}F_{\ell+1}-\frac{\ell}{2\ell-1}F_{\ell-1} &=&0\label{Fnu} \quad  \rm{for} \quad \ell \geq 2~,
\eea
\bea
&&\Phi''(y)+3\HH\eta_1(1+c_s^2)\Phi'(y)+\left[ 3(c_s^2-w)(\HH\eta_1)^2+c_s^2 x_1^2  \right]\Phi(y)=(3c_s^2-1) \frac{( \HH\eta_1)^2 }{2}\, \frac{\Om_\B}{(1+y)^2}\nonumber\\
&&-3w\frac{(\HH\eta_1)^2}{x_1^2}\left\{ \HH\eta_1\Pi_{\rm tot}' (y) +\left[ 2\HH'\eta_1+3\frac{(\HH\eta_1)^2}{w}(w-c_s^2)-\frac{x_1^2}{3} \right] \Pi_{\rm tot}(y)   \right\}~, 
\label{Bardeen}
\eea
where $\Pi_{\rm tot}\equiv R_\nu \pi_\nu+3\Pi_\B $. We differentiate Eq.~(\ref{pinu}) and combine it with Eqs.~(\ref{Dsnu}) to (\ref{Fnu}) and with Einstein's equation~(\ref{Einstein2}), in order to express $\Phi$, $\Phi'$ and $\Phi''$ as functions of $\pi_\nu$ and its derivatives up to fourth order (they would be functions also of $D_{s\,\nu}$ and higher multipoles, but these terms can be neglected since they are at higher order in $x_1\ll 1$). $\Phi$, $\Phi'$ and $\Phi''$ so derived are then inserted back into Bardeen's equation (\ref{Bardeen}): at leading order, this equation becomes then a homogeneous fourth-order differential equation for the variable $\Pi_{\rm tot}$
\be
5y^4\Pi_{\rm tot}^{(4)}+20y^3\Pi_{\rm tot}^{(3)}+8 R_\nu y^2 \Pi_{\rm tot}'' -16 R_\nu y \Pi_{\rm tot}'+16 R_\nu \Pi_{\rm tot}=0~.
\ee 
Solving the above equation, we get
\be
\pi_\nu(y)=-\frac{3}{R_\nu}\Pi_\B+c_1 \frac{\cos(A\log y)}{\sqrt{y}}+c_2 \frac{\sin(A\log y)}{\sqrt{y}}
+c_3 y+c_4 y^2~,
\ee
where $A\equiv \frac{1}{2}\sqrt{\frac{32}{5}R_\nu-1}$. Apart from the terms proportional to $c_3$ and $c_4$, this solution reproduces the one obtained in Eq.~(16) of \cite{Kojima:2009gw}, which has been derived from a second order differential equation for $\pi_\nu$: Eq.~(13) of the same reference. The time dependence of the terms proportional to $c_3$ and $c_4$ would not satisfy the second order differential equation given in \cite{Kojima:2009gw}, which therefore does not admit, strangely enough, the standard inflationary solution for $\pi_\nu$ given for example in \cite{Ma:1995ey}. 

We determine then the constants $c_1$ to $c_4$ by imposing the continuity of $\pi_\nu, V_\nu, D_{s\,\nu}$ and $\zeta$ at decoupling time $y_\nu$, i.e.
\bea
\pi_\nu(y_\nu)&=&0~,\\
\pi_\nu'(y_\nu)&=& \frac{8x_1}{5}V_\nu(y_\nu)~,\\
\pi_\nu''(y_\nu)&=&\frac{8 x_1^2}{5}\left(\Psi+\frac{D_{s\,\nu}}{4} \right)(y_\nu)~,\\
\zeta(y_\nu)&=&\frac{-10a_1}{x_1^2}~,
\eea
where for $\Psi(y_\nu)$ we take our solution $\Psi^{+}(y_\nu)$ given in Eq.~(\ref{Psiplus}) and evaluated well in the radiation era for $y_\nu\ll 1$. Moreover, since we consider adiabatic initial conditions, at leading order in $k\eta_1\ll 1$ we have $V_\nu=V$ and $D_{s\,\nu}=D$. We take therefore for $V_\nu(y_\nu)$ and $D_{s\, \nu}(y_\nu)$ our solutions $V^{+}(y_\nu)$ and $D^+(y_\nu)$ given in Eqs.~(\ref{Vplus}) and (\ref{Dplus}).
We then find for the neutrino anisotropic stress
\bea
&&\pi_\nu(y)=-3\frac{\Pi_\B}{R_\nu}-\frac{40 a_1y^2}{15+4R_\nu}+\left[ \frac{3\Pi_\B}{R_\nu}+\frac{40 a_1y_\nu^2}{15+4R_\nu}\right]\sqrt{\frac{y_\nu}{y}}\cos\left(A\log\frac{y}{y_\nu} \right)\nonumber\\
&&+ \left[\frac{3\big[16 R_\nu(y_\B/y_\nu-1)+5 \big]\Pi_\B}{10 AR_\nu } +\frac{4 a_1(15-16 R_\nu)y_\nu^2}{3(15+4R_\nu)A} \right] \sqrt{\frac{y_\nu}{y}}
\sin\left(A\log\frac{y}{y_\nu} \right)~.
\label{pinusol}
\eea
This solution is plotted in Fig. \ref{fig:Pinu}, section \ref{with_n}. We see that the neutrino anisotropic stress quickly adjusts to the opposite of the magnetic field one, and cancels it. It then starts to grow again, following the standard behaviour $y^2 \sim \eta^2$ coming from the inflationary solution and given for example in \cite{Ma:1995ey}. 

In order to solve the second order equation (\ref{Dddot}) for $D$ analytically after neutrino decoupling, we introduce an approximated function which fits the behaviour in time of Eq.~(\ref{pinusol}), given by
\be
\pi_\nu(y)=\frac{3\Pi_\B}{R_\nu}\left(\frac{y_\nu^2}{y^2}-1 \right)-\frac{40 a_1y(y-y_\nu)}{15+4R_\nu}~.
\ee
This is also plotted in Fig. \ref{fig:Pinu} section \ref{with_n}.

\section{Free-streaming neutrino contribution to the Sachs Wolfe plateau}
\label{app:neutrinosSW}

We evaluate here the contribution of free-streaming neutrinos to the Sachs Wolfe inflationary plateau, due only to their non-zero anisotropic stress (setting the magnetic field to zero). In the standard inflationary case this contribution adds to the purely inflationary result given in \eqref{dToverTinf}. In \eqref{Vgstress}, it is represented by the part of $\Psist-\Phist$ which is proportional to $a_1$, as given in Eqs.~(\ref{Phistress}) and (\ref{Psistress}). In order to evaluate the Sachs Wolfe, we simply have to differentiate \eqref{Vgstress}, as shown in \eqref{SWV}. At leading order in the $x_1$ expansion $\mO(1/x_1^2)$, we find
\bea
\left.\frac{\De T}{T} \right|_{{\rm inf}\,\nu} &=& \Psist_{\rm inf}-\Phist_{\rm inf} \\
&=& -\frac{160 (y_{\rm rec}-y_\nu)\Big[ y_{\rm rec}^3-2y_\nu^2 -2y_{\rm rec}(1+y_\nu)(y_{\rm rec}+y_\nu) \Big] }{3\,y_{\rm rec}^3(2+y_{\rm rec})^3}\frac{R_\nu}{15+4 R_\nu}\frac{a_1}{x_1^2}~. \nonumber
\eea
Setting $y_\nu \simeq 10^{-6}$, $y_{\rm rec} \simeq 1$, $R_\nu \simeq 0.4$ we find
\bea 
&&\left. \frac{\De T}{T} \right|_{\rm inf} \simeq -\frac{2.5\,a_1}{x_1^2} \\
&&\left. \frac{\De T}{T} \right|_{\nu\,{\rm inf}} \simeq \frac{0.05\,a_1}{x_1^2}~, 
\eea
so that the neutrinos reduce the inflationary Sachs Wolfe plateau by about two percent. 

\section{The metric and fluid perturbations at next-to-leading order. }
\label{app:pinunext}

In this appendix we present the derivation of the metric and fluid variables at next-to-leading order, which we compute by solving Eq.~(\ref{phizeta}). We split each variable into its leading order component, which we calculated in Sec.~\ref{with_n} and which we label here with a ${}_0$ subscript, and its next-to-leading order component, which we label with a ${}_1$ subscript.  We use the following Ansatz for the neutrino anisotropic stress
\be
\label{apppinu}
\pi_{\nu 1}(y)=\Big(d_1\Om_B+d_2\Pi_B\Big)(y-y_\nu)^2 x_1^2~,
\ee 
and in order to compute the metric and fluid variables we need to determine the constants $d_1$ and $d_2$. To proceed analytically, we restrict to the radiation dominated era $y\ll1$. We neglect the matter components, and we split the fluid variables into a photon component and a neutrino component
\bea
D&=&(1-R_\nu) D_\gamma+ R_\nu D_\nu ~,\label{eqD}\\
V&=&(1-R_\nu) V_\gamma+R_\nu V_\nu ~.\label{eqV}
\eea
The system of equations for these variables at next-to-leading order is 
\bea
\Phi_1(y)&=&\frac{3}{2}\left(\frac{\HH}{k}\right)^2\Big[(1-R_\nu)D_{\gamma 1}+R_\nu D_{\nu 1} \Big]~,\label{eqphi}\\
\Phi_1(y)+\Psi_1(y)&=&-3w\left(\frac{\HH}{k}\right)^2 R_\nu\, \pi_{\nu 1}~, \label{eqpsi}\\
\Psi_1(y)-\frac{\Phi_1'(y)}{\HH\eta_1}&=&\frac{3}{2}\frac{\HH}{k}(1+w)\Big[(1-R_\nu)V_{\gamma 1}+R_\nu V_{\nu 1}\Big]~, \label{eqv}\\
V'_{\gamma 1}(y)+\HH\eta_1V_{\gamma 1}&=&x_1\Psi_1+\frac{x_1}{4}D_{\gamma 0} + \frac{x_1(\Omega_B-2\Pi_B)}{4(1-R_\nu)}~,\label{eqvg}\\
V'_{\nu 1}(y)+\HH\eta_1V_{\nu 1}&=&x_1\Psi_1+\frac{x_1}{4}D_{\nu 0} - \frac{x_1}{6}\pi_{\nu 0}~,\label{eqvnu}\\
D'_{\gamma 1}(y)-\HH\eta_1D_{\gamma 1}&=&-\frac{4x_1}{3}V_{\gamma 1}~, \label{eqdg}\\
D'_{\nu 1}(y)-\HH\eta_1D_{\nu 1}&=&-\frac{4x_1}{3}V_{\nu 1}-\frac{2}{3}\HH\eta_1\pi_{\nu 1}~, \label{eqdnu}\\
D'_{s\gamma 1}(y)+\frac{4x_1}{3}V_{\gamma 0}&=&-4\Phi'_1~,\label{eqdsg}\\
D'_{s\nu 1}(y)+\frac{4x_1}{3}V_{\nu 0}&=&-4\Phi'_1~,\label{eqdsnu}\\
\pi_{\nu 1}'&=&\frac{8 x_1}{5}V_{\nu 1}~. \label{eqpinu}
\eea
Solving Eq.~(\ref{phizeta}) with the ansatz (\ref{apppinu}), we can express $\Phi_1$ as a function of $d_1$ and $d_2$. Eq.~(\ref{eqpinu}) and Eq.~(\ref{eqpsi}) give $V_{\nu 1}$, respectively $\Psi_1$ as functions of $d_1$ and $d_2$. Inserting these expressions in Eq.~(\ref{eqvnu}) allows to find the constant $d_2$
\be
d_2=\frac{3-2R_\nu}{R_\nu(\rnu)}+\frac{4\log(\frac{y_\B}{y_\nu})}{\rnu}-\frac{3+2d_1(\rnu)}{2(\rnu)}\frac{\Omega_\B}{\Pi_\B}~.
\ee
With this, we find then
\be
\pi_{\nu 1}(y)=\frac{x_1^2(y-y_\nu)^2}{2R_\nu(\rnu)}\left[2(3-2R_\nu)\Pi_\B+8R_\nu\log\left(\frac{y_B}{y_\nu}\right)\Pi_\B-3R_\nu\Omega_\B\right]~.
\ee
We see that $d_1$ and $d_2$ are not independent and that $d_2$ determines completely the solution.

We can then write the full solutions for the metric and fluid variables in the limit $y_B<y_\nu<y\ll 1$, which would correspond, in our approach, to the initial conditions for the Boltzmann hierarchy given in \cite{Finelli:2008xh,PFP,Shaw:2009nf} in the syncronous gauge
\begin{eqnarray*}
\Phi(y)&=&\frac{3(y_B-2y_\nu)}{y^3}\frac{\Pi_B}{x_1^2} +\frac{4(2+R_\nu)\Pi_\B-4(5+2R_\nu)\log\left(\frac{y_\B}{y_\nu}\right)\Pi_\B+R_\nu \Omega_\B}{2(\rnu)}~, \\
\Psi(y)&=&-\frac{3(y_B-2y_\nu)}{y^3}\frac{\Pi_B}{x_1^2} + \frac{-7\Pi_\B+10\log\left(\frac{y_\B}{y_\nu}\right)\Pi_\B+R_\nu\Omega_\B}{\rnu} ~,  \\
D(y)&=&-\Omega_B+\frac{2(y_B-2y_\nu)}{y}\Pi_B+ \frac{4(2+R_\nu)\Pi_\B-4(5+2R_\nu)\log\left(\frac{y_\B}{y_\nu}\right)\Pi_\B+R_\nu \Omega_\B}{3(\rnu)}x_1^2y^2\\
V(y)&=&\left[-\frac{3}{y}-\frac{6(y_\nu-y_B)}{y^2}\right]\frac{\Pi_B}{x_1} +\frac{-7\Pi_\B+10\log\left(\frac{y_\B}{y_\nu}\right)\Pi_\B +R_\nu \Omega_\B}{3(\rnu)}x_1 y\\
D_{s\gamma}(y)&=&\left[\frac{12}{y^2}+\frac{24(y_\nu-y_B)}{y^3}\right]\frac{\Pi_B}{x_1^2} +  \frac{2(10 R_\nu-29)\Pi_\B+40(1-R_\nu)\log\left(\frac{y_\B}{y_\nu}\right)\Pi_\B  +\Big( 30-(3+8R_\nu)R_\nu \Big)\Omega_\B }{2(R_\nu-1)(\rnu)}~,\\
V_\gamma(y)&=&\left[-\frac{3}{y}-\frac{6(y_\nu-y_B)}{y^2}\right]\frac{\Pi_B}{x_1} + \frac{2(29-10R_\nu)\Pi_\B+40(R_\nu-1)\log\left(\frac{y_\B}{y_\nu}\right)\Pi_\B-19R_\nu\Omega_\B}{8(R_\nu-1)(\rnu)}x_1 y~,\\
D_{s \nu}(y)&=&\left[\frac{12}{y^2}+\frac{24(y_\nu-y_B)}{y^3}\right]\frac{\Pi_B}{x_1^2}  + \frac{10(2R_\nu-3)\Pi_\B-40R_\nu\log\left(\frac{y_\B}{y_\nu}\right)\Pi_\B-R_\nu(15+8R_\nu)\Om_\B}{2R_\nu(\rnu)}~,\\
V_\nu(y)&=&\left[-\frac{3}{y}-\frac{6(y_\nu-y_B)}{y^2}\right]\frac{\Pi_B}{x_1}  + \frac{10(3-2R_\nu)\Pi_\B+40 R_\nu\log\left(\frac{y_\B}{y_\nu}\right)\Pi_\B-15R_\nu\Om_\B}{8R_\nu(\rnu)}x_1 y~,\\
\pi_\nu(y)&=&3\left[ -1+\frac{y_\nu^2}{y^2} \right] \frac{\Pi_B}{R_\nu}+ \frac{2(3-2R_\nu)\Pi_\B+8R_\nu\log\left(\frac{y_\B}{y_\nu}\right)\Pi_\B-3R_\nu\Om_\B}{2R_\nu(\rnu)} x_1^2 y^2~.
\end{eqnarray*}

\end{document}